# Knowledge-Certainty Diversity Predicts Human-Human and Human-AI Open-Ended Discussion Synergy and Reveals Pure-AI Discussion Shortfalls

Tom Sheffer[1,2], Alon Miron[1], Asael Sklar[3], Yaniv Dover[1], Ariel Goldstein[1,2,4]

[1] The Hebrew University of Jerusalem, Jerusalem, Israel [2] Google Research, Tel Aviv, Israel [3] Reichman University, Herzliya, Israel [4] The Centre for Human-Inspired Artificial Intelligence, University of Cambridge, United Kingdom

Corresponding author: Tom Sheffer | tom.sheffer@mail.huji.ac.il

## Abstract

Conversations transform individual knowledge into collective insight, enabling collaborators to solve problems more accurately than they could alone. Whether dialogues among large language models (LLMs) can replicate the synergistic gains observed in human discussion remains unclear. We systematically compared four interaction settings, LLM–LLM pairs, LLM trios, human trios, and human–LLM pairs, using validated medical multiple-choice questions. Agents answered individually, engaged in open-ended discussion, then re-answered, allowing us to quantify conversational gains. Interactions that included humans consistently yielded synergy (post-discussion accuracy increased for both stronger and weaker participants), whereas purely LLM groups did not improve and often declined. To explain and anticipate when open-ended dialogue helps, we introduce an agent-agnostic knowledge-certainty framework that models each participant by performance (accuracy) and confidence. This framework quantifies knowledge-certainty diversity, the degree to which one agent tends to be correct when another is uncertain, and yields an upper bound on the gain available from choosing between the agents' answers on the basis of confidence, which we term Potential Conversation Synergy. Computed before any discussion, this metric was low in both settings where discussion did not help and high in both where it did: When knowledge-certainty diversity is low (as in LLM-only groups), discussion doesn't improve performance; when it is present (as in human or human–LLM groups), open-ended dialogue reliably lifts accuracy. These findings suggest a diagnostic for AI collaboration: knowledge-certainty diversity can be measured from pre-discussion responses and provides a configuration-level indicator of conversational gains across the settings studied.

## Introduction

Conversation is a fundamental human faculty for knowledge enhancement and collaborative problem-solving [1–3]. Through dialogue, individuals effectively pool diverse insights and perspectives, enabling collective outcomes that surpass isolated individual efforts [4–6]. This collaborative phenomenon has been widely recognized as a key driver of successful group decision-making and problem-solving across a range of domains [7]. Complementary work frames human conversation as the engine for surfacing and integrating informational diversity: nonredundant knowledge, cues, and perspectives distributed across members. Across these accounts, gains tend to be greatest when groups actively elaborate complementary expertise rather than merely aggregating opinions [8,9].

Inspired by this human ability, numerous implementations of interactions between LLM agents have recently emerged, including collaborative human–AI dialogues and fully automated multi-agent simulations [10–13]. In parallel, an increasingly active branch of research seeks to define the conditions under which human–AI collaboration succeeds: LLM-powered chatbots can facilitate complex clinical workflows such as primary-to-specialist care transitions [14], tailoring the human-AI interaction context to individuals' cognitive styles can improve performance on unstructured managerial tasks[15], and work on the "human–algorithm connection" emphasizes the importance of teaming dynamics [14,16]. Yet it remains unclear whether interactions involving LLM agents genuinely replicate the synergistic properties observed in fully human conversations or simply mimic dialogue without achieving comparable collaborative depth [17,18].

A related line of work asks how AI can improve a group's outcome. Some systems mediate: the Habermas Machine elicits opinions privately and drafts statements that win broad endorsement, including from the minority [19], building on earlier work generating consensus statements for groups with diverse views [20]. Others facilitate, suggesting rephrasings that improve conversation quality without shifting either side's position [21]. Both address questions with no correct answer. Where correctness is defined, small deliberating groups can aggregate knowledge better than much larger crowds voting alone [22]. Our agents converse directly, as peers, on questions with a correct answer, and no third party shapes the exchange. We examine whether confidence-conditioned complementarity measured before discussion is associated with conversational gains across the configurations studied.

Another line of work demonstrates that aggregation of agents' responses can improve accuracy in classical ensembles and in engineered multi-agent pipelines that impose roles, search, or aggregation protocols [23–26]. However, these methods focus on structured procedures designed to optimize performance and do not provide a prospective diagnostic of when *open-ended* dialogue will help. Here, we take a complementary approach: rather than engineering coordination, we ask when open-ended conversation improves accuracy, and we generalize the answer across humans, LLMs, and mixed teams.

To systematically investigate this question and construct a diagnostic measure, we assessed conversational interactions across Human-Human, Human-LLM, and LLM-LLM interactions within a medically oriented setting. We selected the medical domain specifically because medical reasoning inherently involves extensive knowledge exchange and rich conversational interactions [27,28], providing a realistic and challenging environment to evaluate conversational synergy. This domain also carries significant societal importance. Effective AI integration could improve healthcare access in resource-limited settings and help accelerate critical innovations such as drug discovery [29,30]. Notably, the medical field has already attracted considerable practical interest and progress in LLMs. Systems like AMIE [31] have demonstrated clinical and diagnostic capabilities comparable to, and in some cases exceeding, those of physicians, suggesting the potential for integrating such models into clinical teams.

A related line of work quantifies what single models “know”: factual coverage, retrieval, and confidence calibration [32,33]. Here we shift from defining knowledge as performance and include confidence as a fundamental dimension in modeling the group’s potential for increased performance as a result of open-ended deliberation. Building on cognitive theories of human reasoning [34] and recent findings on language-model confidence [35,36], we introduce an agent-agnostic framework that represents each participant, human or LLM, by two quantities: performance (i.e., accuracy) and confidence. We refer to this joint profile as an agent's knowledge-certainty: what it answers correctly, together with how certain it is when answering. The two dimensions are distinct, and the framework depends on their combination, since an agent can be accurate but unsure, or certain but wrong. Knowledge-certainty diversity is present when one agent tends to be correct on the items where the other agent reports low confidence. This is a narrower condition than disagreement. Error-diversity measures from ensemble learning count how often agents disagree [37], but disagreement alone does not tell either agent whose answer to keep. Confidence calibration [35,36] tells an agent when its own answer is unreliable, but says nothing about whether a partner can do better. Confidence-weighted aggregation and self-consistency [38,39] combine answers that agents commit to independently, which requires a central aggregator and cannot use anything said in a conversation. Knowledge-certainty diversity requires both parts together: an agent that knows when it is uncertain, and a partner that is more often correct precisely at those points.

Potential Conversation Synergy (PCS, formally defined below) measures this. It is the accuracy a group would gain if each agent deferred to its partner exactly at the confidence levels where the partner is more often correct, computed from pre-discussion answers and confidence. The underlying rule is a form of confidence-based deferral, studied in human pairs [40] and in work on human–AI complementarity and deferral [41–43], where a system applies the rule on the agents' behalf. Here we do not apply it. Instead, we ask whether a group that is simply left to converse reaches the gain the rule predicts. This question is open: human–AI combinations often fail to beat their best member [44], and no existing measure says in advance which groups will benefit from discussion.

Our results indicate that purely LLM-based discussions fail to improve accuracy, whereas conversations involving humans exhibit notable synergy, collectively surpassing the performance of the best individual agent (human or LLM) prior to interaction, and yield measurable performance gains. Our experiments revealed that human agents naturally exhibit substantial knowledge-certainty diversity and effectively utilized their internal confidence to strategically enhance collective performance, surpassing even the strongest individual's initial performances. In contrast, pure LLM interactions showed no such improvement, an outcome anticipated by, and consistent with, their low knowledge-certainty diversity. However, when paired with humans, LLMs achieved significant accuracy gains. Although large language models can be sensitive to prompt wording and format [45,46], replications across decoding temperatures and a second model set (Figs. S1–S4, S8) argue against a narrow decoding artefact. However, we did not test alternative discussion protocols and prompt-independence remains a hypothesis. Whenever knowledge-certainty diversity was present, whether in human-human or human-LLM conversations, our open-ended setup yielded immediate gains without assigned epistemic roles or imposed aggregation rules.

## Results

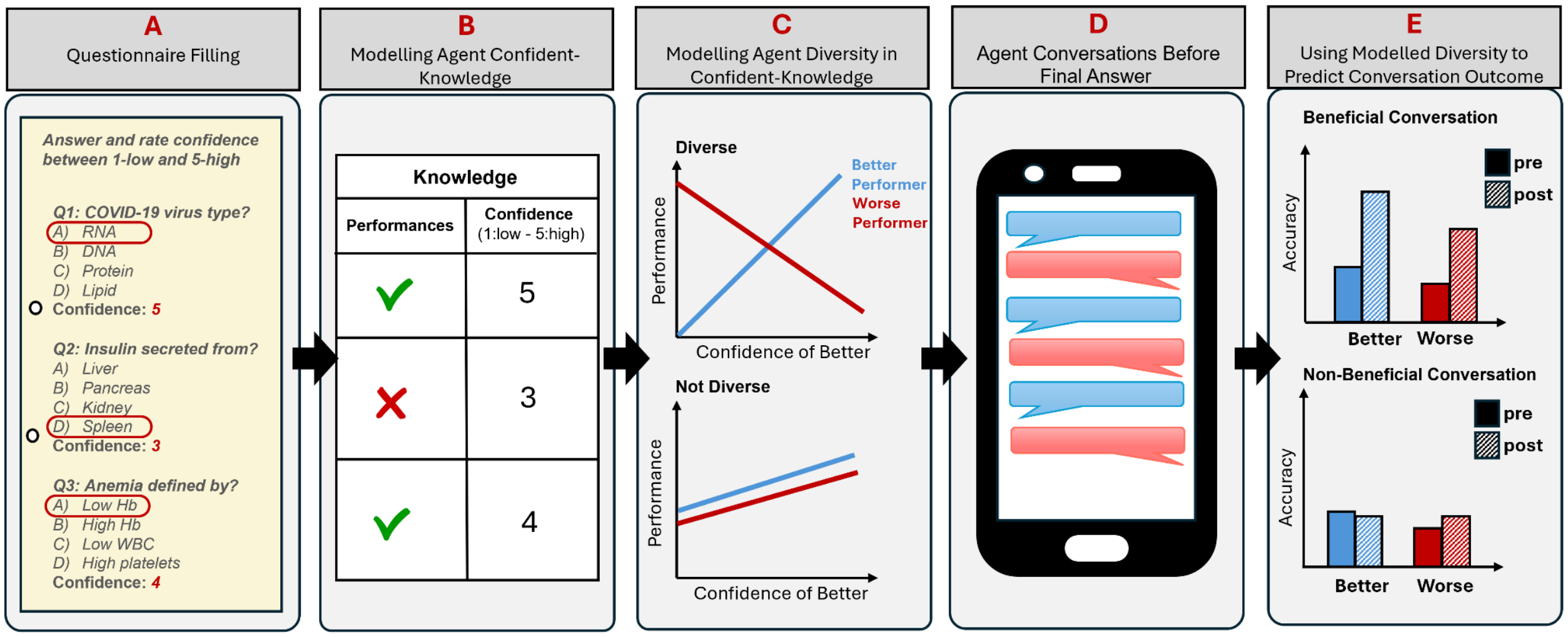

__Figure 1.__ **Overview of experimental pipeline and diversity framework. A. Questionnaire Filling.** Each agent (human or artificial) answers a block of multiple-choice medical questions. Human agent's confidence was asked explicitly and LLM agent's confidence was evaluated using self-consistency repetitions. **B. Modeling Agents' knowledge-certainty.** An agent's initial knowledge-certainty is represented by two dimensions—accuracy (correct vs. incorrect answers) and confidence (human self-report: 1–5; LLM self-consistency: 1–5 in Experiments 1–2 and 0–5 in Experiment 4). **C. Modeling knowledge-certainty  Diversity.** The schematic shows how accuracy for the better and worse agents varies with the better agent's confidence. In the top panel the curves cross: when the better agent is unsure, the worse agent is more accurate, exemplifying high knowledge-certainty  diversity. In the bottom panel the curves never meet, representing low diversity. **D. Agent Conversations.** Agents engage in open-ended, text-based discussion before submitting final answers. **E. Outcomes.** When diversity is high (top bars), both better and worse agents improve after conversation. When diversity is low (bottom bars), little or no improvement is observed.

We examined four conversational configurations - LLM-LLM pairs (Experiment 1), LLM trios (Experiment 2), human trios (Experiment 3), and human-LLM pairs (Experiment 4; Table 1) — each tested using validated medical multiple-choice questions covering core biomedical facts and clinical vignettes in the style of licensing examinations routinely used to evaluate clinicians and senior medical students. The human participants were medical students in their clinical years, specifically selected due to their specialized expertise in solving medical multiple-choice questions as preparation for national licensing examinations, a requirement that naturally limited the available participant pool. Agents (humans or artificial) first answered questions individually, then engaged in open-ended discussions, and finally re-answered the same questions, allowing us to quantify how discussion altered performance (see Fig. 1 & Materials and Methods). Humans explicitly rated their confidence on a 1–5 scale after each question. For LLMs, we estimated confidence via self-consistency: five independent runs in Experiments 1–2 and ten in Experiment 4, with confidence defined relative to the model's pre-discussion answer and mapped to the scales described in Methods, which prior work shows is calibrated and internally consistent [35,47] (Fig. 1A). Combining accuracy and confidence yielded a quantitative knowledge-certainty profile for each agent (Fig. 1B). We use the performance and confidence of the initial answers to estimate Potential Conversation Synergy which we then compare with the conversational gain (Fig. 1C). After submitting their initial answers, agents then entered an open-ended, text-based discussion (Fig. 1D). They then re-answered the same questions, and we quantified the resulting performance changes to evaluate the impact of the conversation (Fig. 1E).

| | **Exp 1: LLM–LLM pairs** | **Exp 2: LLM trio** | **Exp 3: Human trios** | **Exp 4: Human–LLM pairs** |
|---|---|---|---|---|
| **Agents** | Claude Opus 4, OpenAI o3, Gemini 2.5 Pro (3 pairs) | The same three models, one trio | Medical students, 4th or 6th year; 30 in 10 discussion trios, 30 no-discussion controls | 32 medical students recruited, 30 with complete paired records, each paired with GPT-4 Turbo |
| **Items** | 200: 50 each from MedQA-USMLE, MedMCQA, MedBullets, MedXpertQA | The same 200 | 42 internal-medicine items, in-house, clinician-vetted | 50 paediatric items, in-house, clinician-vetted |
| **Answer format** | 4–10 options, order shuffled | 4–10 options, order shuffled | Binary | 4 options |
| **Language** | English | English | Hebrew | Hebrew discussion; English question |
| | | | | inputs for the LLM |
| **Confidence** | 5 self-consistency samples; level is the number agreeing with the modal answer (1–5) | Same | Self-report, 1–5 | Human self-report 1–5; LLM 10 self-consistency samples, halved and rounded up (0–5) |
| **Discussion protocol** | Fixed debate prompt, randomised speaking order, 6 messages, then re-answer | Same, 6 messages | Synchronous group chat, timed, consensus not required | Open-ended chat, participant-paced |
| **Item selection** | All items | All items | All items | Participant-selected: 562 of 1,550 instances (36%) |
| **Unit of inference** | 12 model pair–benchmark combinations | One configuration; no test | 27 participants / 10 trios | 30 participant–LLM pairs |

**Table 1. Experimental designs across the four settings**. The experiments differ in question sets, answer format, language, confidence measurement, interaction protocol, and item selection. Within-setting analyses describe outcomes under each design. Cross-setting comparisons are descriptive and do not isolate the effect of team composition.

## Potential Conversation Synergy

At the core of our analysis is knowledge-certainty diversity, which we define as the extent to which agents' knowledge is complementary. A group possesses high knowledge-certainty diversity when one agent's low-confidence cases systematically coincide with a partner's higher likelihood of being correct. We visualize knowledge-certainty diversity by plotting each agent's pre-discussion accuracy against the best performer's confidence level, revealing whether interaction enables the best performer, and by extension the group, to exceed the best performer's initial accuracy (Fig. 1C). If, at some confidence levels of one agent, the other agent is more often correct, the group has knowledge-certainty diversity (Fig. 1C top). We demonstrate that knowledge-certainty diverse groups' discussions improve their performance. Conversely, if the partner is not more accurate at any confidence level (Fig. 1C bottom), or is more accurate only at levels containing very few items, knowledge-certainty diversity remains limited even when agents excel on different subsets of items.

We also quantify collaborative potential with a metric we term Potential Conversation Synergy. Potential Conversation Synergy measures the improvement a group would obtain if the reference agent adopted its partner's answer at exactly those confidence levels at which the partner is more accurate (Methods). Recalculating accuracy under those choices gives the Potential Conversation Synergy. Its values are expressed in percentage-point improvement and range from 0 (no additional improvement possible) up to 100 minus the best agent's starting accuracy. For example, if the top-performing agent begins at 80%, the maximum attainable Potential Conversation Synergy is 20 percentage points. Low values indicate that confidence-aware collaboration offers little room for improvement, whereas high values reveal substantial complementary knowledge-certainty that could be exploited through conversation.

Potential Conversation Synergy is thus a simplified diversity model anchored in a single agent's confidence: the better performer's in pairs, the dissenter's in trios. It is an upper bound on gains from confidence-based selection between the two initial answers, not on conversational gains in general (Methods). We also report a multi-confidence variant (Fig. S5) in which the oracle conditions on all agents' confidence for each item (not just of one teammate). This richer model preserves the qualitative pattern across settings and leaves our conclusions unchanged.

We include confidence signals in modeling Potential Conversation Synergy due to two theoretically and empirically supported properties observed in both humans and modern LLMs. First, agents display calibrated confidence: higher self-rated (for humans) or self-consistency (for LLMs) confidence scores reliably predict greater accuracy (Fig. S6). Second, both types of agents behave as confidence-sensitive switchers: in our data, both humans and language models switched answers more often when their own confidence was low (Fig. S7), and this confidence-tracked switching is consistent with the gains we observe. These two findings provide the foundation for our framework. Synergy was therefore not unpredictable: it was largest where one agent's confidence on a question was low and its partner was more often correct at that level.

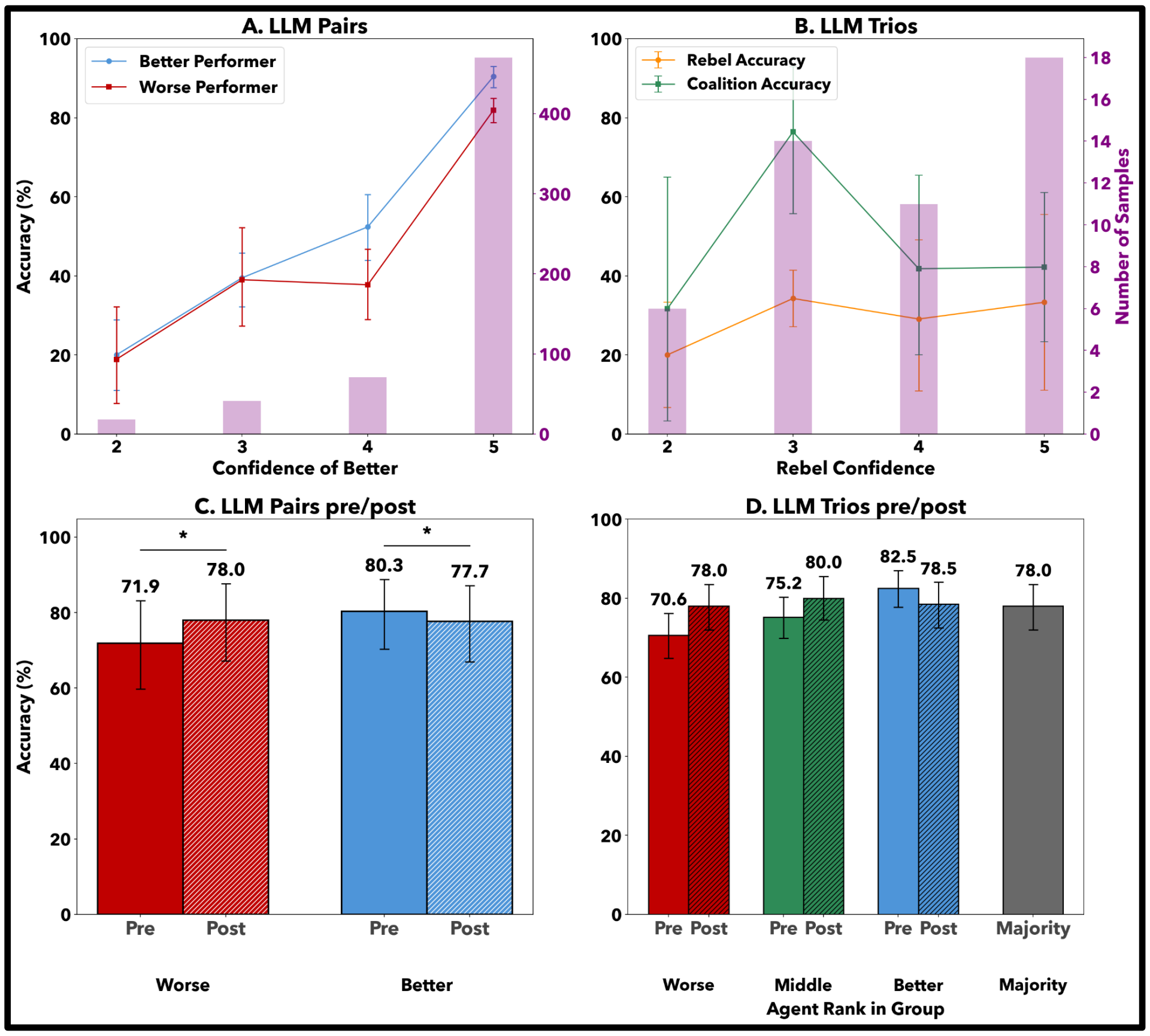

***<u>Figure 2.</u> LLM-only interactions: Pairs and Trios.***

**A. Knowledge-certainty diversity in LLM-LLM pairs (Experiment 1; 3 model pairs, 600 pair-items).** Accuracy of the better (blue) and worse (red) agents is plotted against the better agent's questions split on their confidence level. No meaningful crossover is observed. The worse agent is not more accurate at any level (levels 2–4: two-sided exact paired sign-test p = 0.55, 1.00, 0.21); PCS = 0.0 pp (95% CI 0.0–0.6; Table 2). **B. Knowledge-certainty diversity in LLM trios (Experiment 2; 49 items with a 2–1 split)**. Rebel (orange) and coalition (green) accuracy is plotted by rebel confidence. No level shows a crossover; the coalition is nominally ahead at every level; PCS = 0.0 pp (95% CI 0.0–2.9; Table 2). **C. Pre- and post-conversation accuracy for LLM-LLM pairs** for the initially worse (red) and better (blue) agents. The worse agent's accuracy rises from 71.9% to 78.0% (paired t(11) = 3.02, p = 0.012); the better agent's falls from 80.3% to 77.7% (t(11) = −2.66, p = 0.022). **D. Pre- and post-conversation accuracy for LLM trios** by initial rank, with the pre-discussion majority vote (grey). After discussion, all three models are close to the majority vote (78.0%; worst 78.0%, middle 80.0%, best 78.5%); the best model fell from 82.5% to 78.5%. Error bars are 95% percentile bootstrap intervals, resampling items within confidence levels (A, B), the 12 model pair–benchmark combinations and recomputing pooled accuracy (C), or the 200 items in the single trio configuration (D). Purple bars show item counts per confidence level.

**Low Potential Conversation Synergy in Pure LLM Groups Predicts Lack of Conversational Gains**

Analyzing the knowledge-certainty diversity between each pair of the three current state-of-the-art large language models: OpenAI o3, Google Gemini 2.5 Pro, and Anthropic Claude Opus 4 revealed low conversational potential. Even though there is variance in their performance (70.6%, 75.2% and 82.5%; SD = 6.0 pp) when evaluating knowledge-certainty diversity in these pairs we see that they are not diverse and thus have low Potential Conversation Synergy (PCS). We plotted each agent's accuracy as a function of the initially better agent's confidence levels (Fig. 2A). At no confidence level was the worse agent more accurate than the better agent, so Potential Conversation Synergy was 0.0 pp. The multi-confidence variant that conditions on both agents' confidence yielded a similarly small estimate (0.5 pp; Fig. S5). Such a marked lack of knowledge-certainty diversity, constrained their potential to collaboratively improve accuracy.

This low Potential Conversation Synergy matches their consistent failure to enhance group accuracy during conversation (Experiment 1; Fig. 2C). Indeed, the initially better-performing agent, defined by higher pre-discussion accuracy, did not show any improvement in post-conversational accuracy. The better-performing agent's accuracy declined from 80.3% to 77.7% (paired $t(11) = -2.66$, $p = 0.022$), while the initially worse-performing agent rose from 71.9% to 78.0% ($t(11) = 3.02$, $p = 0.012$). We replicated the analysis with a second set of widely used, mid-tier LLMs on an extended question pool, finding the same no-gain pattern (Fig. S8).

To explore conversational dynamics beyond pairs, we extended our analyses to trios. This trio configuration enabled us to compare performance against the established baseline of majority-vote aggregation [6] and to assess knowledge-certainty diversity in multi-agent interactions. To measure knowledge-certainty diversity within these trios, we developed a structured framework for analyzing disagreements. Questions were the same multiple-choice items used in Experiment 1. On items where exactly one agent dissented from the other two, we designated that agent the 'rebel' and the other two the 'coalition'. On items where all three agents gave different answers (8 out of the 200) there is no coalition; as with unanimous items, these remain in the item count and contribute no gain. To explicitly model, visualize and calculate Potential Conversation Synergy and its potential role in collaborative improvement, we plotted rebel accuracy alongside coalition accuracy as a function of the rebel’s internal confidence. This approach allowed us to evaluate whether rebel agents, guided by their confidence, could strategically adopt or resist the coalition's opinion. Such behavior would indicate potential to enhance overall group performance. In the LLM trios, however, rebels never significantly outperformed the coalition at any level of confidence resulting in Potential Conversation Synergy of 0.0 pp (Fig. 2B). The multi-confidence variant that conditions on both the rebel’s and coalition’s confidence raises the value to 1 pp (Fig. S5), which remains small. The absence of a meaningful crossover point suggests a fundamental lack of complementary knowledge among these models predicting low potential for knowledge-certainty diversity-driven improvements.

Aligned with that finding, LLM trios (Experiment 2) did not demonstrate overall accuracy gain following discussions across 200 questions (Fig. 2D): the best-performing model’s accuracy decreased from 82.5% to 78.5%, the middle-ranking model’s accuracy increased from 75.2% to 80.0%, and the lowest-performing model increased from 70.6% to 78.0%.

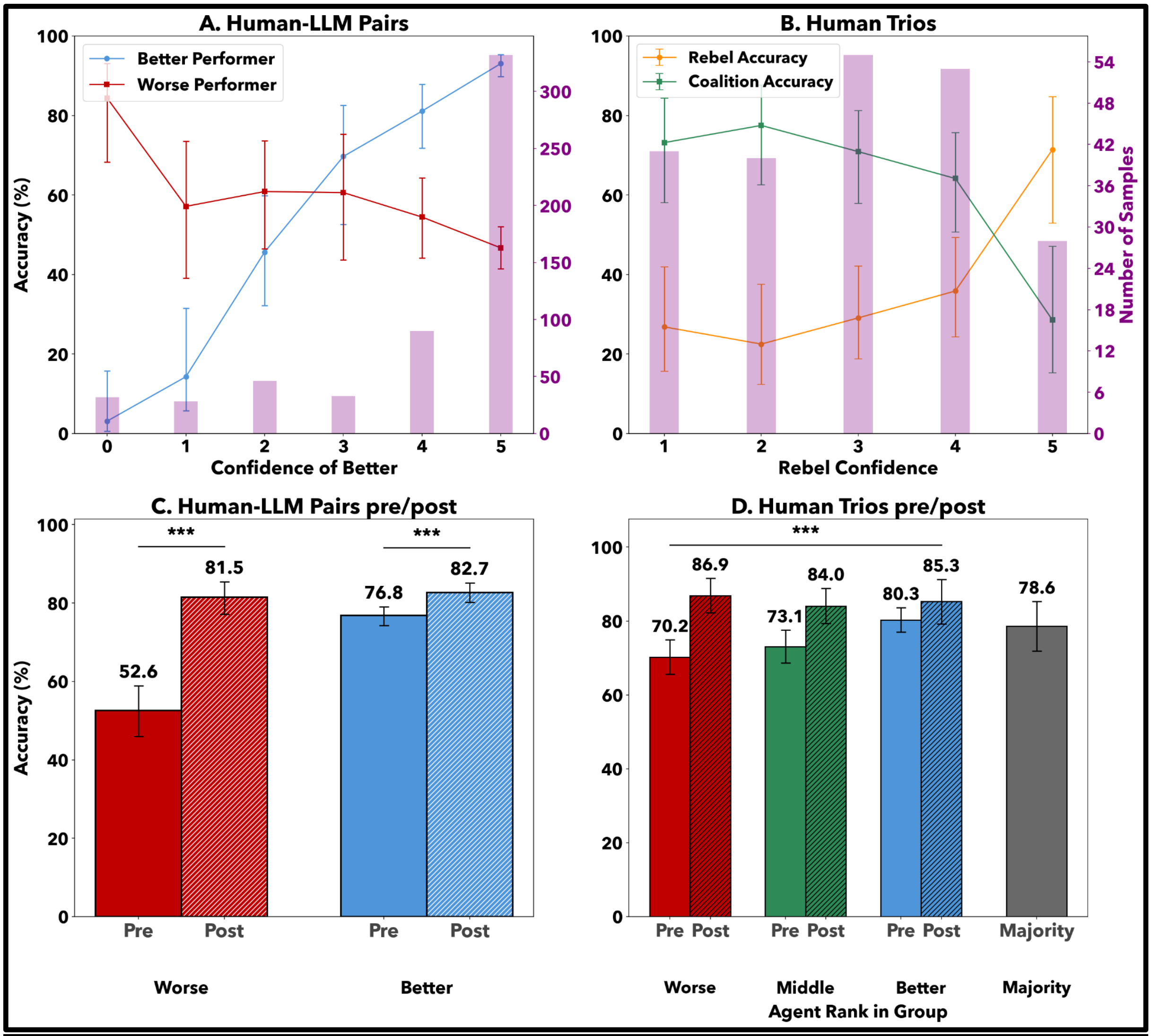

***Figure 3. Human-involving interactions: Trios and Human-LLM Pairs.***

**A. Knowledge-certainty diversity in human-LLM pairs** (Experiment 4; 30 pairs, 561 analyzed items). Accuracy of the better (blue) and worse (red) agents is plotted by the better agent's confidence. Confidence is the reference agent's: the language model's in 26 of 30 pairs and the participant's in the remaining four (Methods). At confidence 0–2 (106 items, 19%) the worse partner is the more accurate (67.0% vs. 24.5%; two-sided exact paired test, $p < 0.001$); PCS = 8.0 pp (95% CI 5.9–10.3; Table 2). **B. Knowledge-certainty diversity in human trios** (Experiment 3; 10 trios, 217 trio-items with a 2–1 split). Rebel (orange) and coalition (green) accuracy by rebel confidence. The coalition is more accurate at rebel confidence 1–3 ($p \leq 0.004$) and, nominally, at 4 ($p = 0.053$); confident rebels are more accurate at level 5 (28 items; 71.4% vs. 28.6%, $p = 0.036$); PCS = 2.9 pp (95% CI 0.5–5.5; Table 2). **C. Pre- and post-conversation accuracy for human-LLM pairs** for the initially worse (red) and better (blue) agents, both improve: 52.6% to 81.5% (paired $t(29) = 8.4$) and 76.8% to 82.7% ($t(29) = 4.5$), both $p < 0.001$. **D. Pre- and post-conversation accuracy for human-only trios** by initial rank ($n = 6$, 10, 11), with the pre-discussion majority vote (grey; 10 trios). Pooled, accuracy rises from 75.4% to 85.2% (paired $t(26) = 6.48$, $p < 0.001$; bracket); best-ranked participants rise from 80.3% to 85.3% (t(10) = 2.91, p = 0.016), above the majority vote (78.6%). Error bars are 95% Wilson intervals (A, B), 95% percentile bootstrap intervals for pooled accuracy obtained by resampling the 30 participant–LLM pairs (C), and 95% t-based CIs across participants within each rank (D). Purple bars show item counts per confidence level.

Because decoding temperature can affect both our confidence estimates and conversational outcomes, we replicated all analyses at additional temperatures. Confidence calibration held (Fig. S1), item-level confidence ranks were stable across temperatures (Fig. S2), and Potential Conversation Synergy was 0.43 and 0.00 pp for pairs and 2.1 and 0.0 pp for trios at temperatures 0.2 and 0.5, respectively; neither temperature yielded a statistically significant better-agent gain (Figs. S3–S4).

**Potential Conversation Synergy in Human-Involving Groups Predicts Conversational Gain**

In sharp contrast, human trios (Experiment 3) naturally demonstrated knowledge-certainty diversity. We used the same knowledge-certainty diversity modeling for trios to explore the level of Potential Conversation Synergy among the human trios (Fig. 3B). This analysis showed that when rebels in human trios were highly confident (level 5) and disagreed with the coalition, they were right more often: on 20 of these 28 items, against 8 for the coalition (71.4% vs. 28.6%; p = 0.036; Fig. 3B). These items made up 7% of trio-items (28 of 420) and resulted in a Potential Conversation Synergy of 2.9 pp. The multi-confidence variant that uses both the rebel's and coalition's confidence increases the value to 5.7 pp (Fig. S5). This suggests that in such cases, rebels were not simply disagreeing at random but often held accurate, complementary knowledge absent from the majority. Importantly, their high internal confidence provided a valuable signal indicating the likely accuracy of their answers, demonstrating that internal confidence effectively identifies valuable minority insights and can significantly enhance group decision-making beyond simple majority rule. In line with this high Potential Conversation Synergy, we found that collaboration led to significant performance gains.

In our experiment, participants were randomly assigned to either a discussion (intervention) or no-discussion (control) condition. In the discussion group (N=30), overall participant accuracy significantly increased (Fig. 3D) from 75.4% (SD = 6.7) to 85.2% (SD = 7.15) (paired t-test: $t(26) = 6.48$, $p < 0.001$). The accuracy of the initially best-performing participant improved from 80.3% (SD = 4.9) to 85.3% (SD = 9.0) (paired t-test: $t(10) = 2.91$, $p = 0.016$), significantly exceeding the pre-discussion majority vote accuracy of 78.6% (paired t-test: $t(10) = 2.52$, $p < 0.05$). As a simple aggregation baseline, we computed the pre-discussion majority vote for both trio settings. In both human and LLM trios this baseline was below the best individual on average. In LLM trios, majority-vote accuracy changed little after discussion (78.0% to 78.5%; Table 2). In human trios, where Potential Conversation Synergy indicated complementary knowledge, post-discussion scores exceeded the pre-discussion majority (Fig. 3D), consistent with the gains reported above.

The no-discussion control condition (N=30) controlled for the effect of mere repetition. Control-group participants showed no significant improvement, confirming that accuracy gains in the intervention condition arose specifically from discussion rather than repeated exposure to questions. Overall participant accuracy in the control condition remained stable from 76.6% (SD = 8.3) to 77.3% (SD = 7.6). The best-performing control participants showed a slight decline, from 83.3% to 82.1% (paired t-test: $t(9) = 1.00$, $p = 0.34$). Direct comparison of score changes between the intervention and control conditions clearly illustrates the advantage of conversation: overall participant improvement was significantly higher in the intervention group (9.79%) compared to the control group (0.73%) (t-test: $t(51) = 5.17$, $p < 0.001$). Likewise, improvement of the best-performing participants was significantly greater in the intervention group (+4.98%) compared to a decrease in the control group (-1.19%) (t-test: $t(19) = 2.90$, $p = 0.009$).

**Introducing knowledge-certainty Diversity to LLM Conversations via Human Interaction**

Recognizing the difficulty of naturally introducing knowledge-certainty diversity among LLM-only groups, we paired human participants with LLMs (using GPT-4 Turbo) to inject knowledge-certainty diversity into these interactions (Experiment 4). We analyzed 562 four-option questions from human-LLM conversations (N=32). Initially, participants individually answered medical questions, subsequently selecting which questions to discuss further with the LLM. To eliminate benchmark contamination, we evaluated this experiment on a new, contamination-free set of translated, locally administered pediatric board items vetted by a senior physician, which being harder and non-public yield lower raw accuracies than standard public 4-option benchmarks. Humans freely decided the number and choice of questions for discussion before re-answering all, averaging 18 discussed questions per participant (Total questions = 50, SD = 9). Using the previously discussed Potential Conversation Synergy framework, we found that in those interactions, for approximately 20% of questions where the better agent's confidence was low (level 2 or below), their accuracy fell below that of the less confident partner (Fig. 3A). This resulted in a Potential Conversation Synergy of 8.0 pp; In the multi-confidence variant, which uses both agents' confidence on each item, the value increases to 10.7 pp (Fig. S5), underscoring a larger potential for collaborative gain between the agents. Under these conditions, the more confident and accurate responses provided by the worse performing partner allowed the better agent to beneficially revise its answers. Additionally, clear knowledge-certainty diversity was also revealed when analyzing the interaction from the perspective of the initially worse-performing agent illustrating how lower-performing agents can effectively use both dialogue and internal confidence signals to achieve substantial performance gains (Fig. S9).

Consistent with this, Human-LLM pairs consistently improved accuracy through discussion (Fig. 3C). Accuracy for the initially better-performing agent increased from 76.8% (SD = 8.6) to 82.7% (SD = 9.4; paired t-test: $t(29) = 4.5$, $p < 0.001$). The initially worse-performing agent showed a greater improvement, rising from 52.6% (SD = 18.6) to 81.5% (SD = 15.1; paired t-test: $t(29) = 8.4$, $p < 0.001$). Post-discussion, both agents surpassed the accuracy initially demonstrated by the better-performing agent, highlighting substantial mutual benefits from human-LLM interactions.

The 1,550 recorded participant–question instances comprised 214 with no LLM response, 774 with an LLM answer shown without a qualifying discussion, and 562 analyzed discussions. Initial participant confidence was lower for discussed instances than for answer-shown or no-response instances (2.86, 3.13, and 4.22, respectively), as was participant accuracy (53.7%, 76.7%, and 89.7%). Discussed questions also had lower mean cohort accuracy (64.9%, 71.9%, and 78.0%; all discussion-versus-comparison $p \leq 0.006$, with participant-clustered standard errors; Supplementary Table S10). Among instances with valid recorded model answers, initial LLM accuracy was lower in the discussion group than in the answer-shown comparison group (75.3% versus 90.1%), and initial disagreement was higher (63.9% versus 23.7%; both $p < 0.001$). These differences characterize the selected discussion subset but do not eliminate selection bias. The conversational gains reported here therefore apply to the analyzed discussions.

Among the 774 instances with an LLM answer shown without a qualifying discussion, participant accuracy increased from 76.7% to 90.8%. Among the 214 instances with no LLM response, aggregate participant accuracy remained at 89.7%. These groups were not randomly assigned, so the comparisons do not isolate the causal effects of exposure or discussion.

Further analyses showed that in most conversations (26 out of 30; two participants chose not to discuss any questions with the LLM), GPT-4 Turbo began as the better-performing agent, with an average initial accuracy of 75.8%, compared to 53.7% for the human participants. Even when evaluating only the interactions in which GPT-4 Turbo encountered a less knowledgeable partner, the model still consistently improved its performance. Its accuracy increased from 77.4% ($SD = 8.8$) to 83.0% ($SD = 8.3$), a statistically significant gain (paired t-test, $t(25) = 4.21$, $p < 0.001$). This demonstrates the model's capacity to extract useful information even from partners with lower overall accuracy.

We further examined how agents leveraged their internal confidence signals to guide decision-making during disagreements. In low-confidence cases (levels 0-2) where initial disagreement occurred, the initially better agent eventually adopted the partner's answer in 47.7% of instances. In contrast, in high-confidence cases (levels 3-5), such changes occurred significantly less, in only 8.9% of disagreements (Z-test: $Z = 8.09$, $p < 0.001$). This finding highlights how internal confidence facilitated beneficial answer switching, consistent with confidence-based switching contributing to the observed gains.

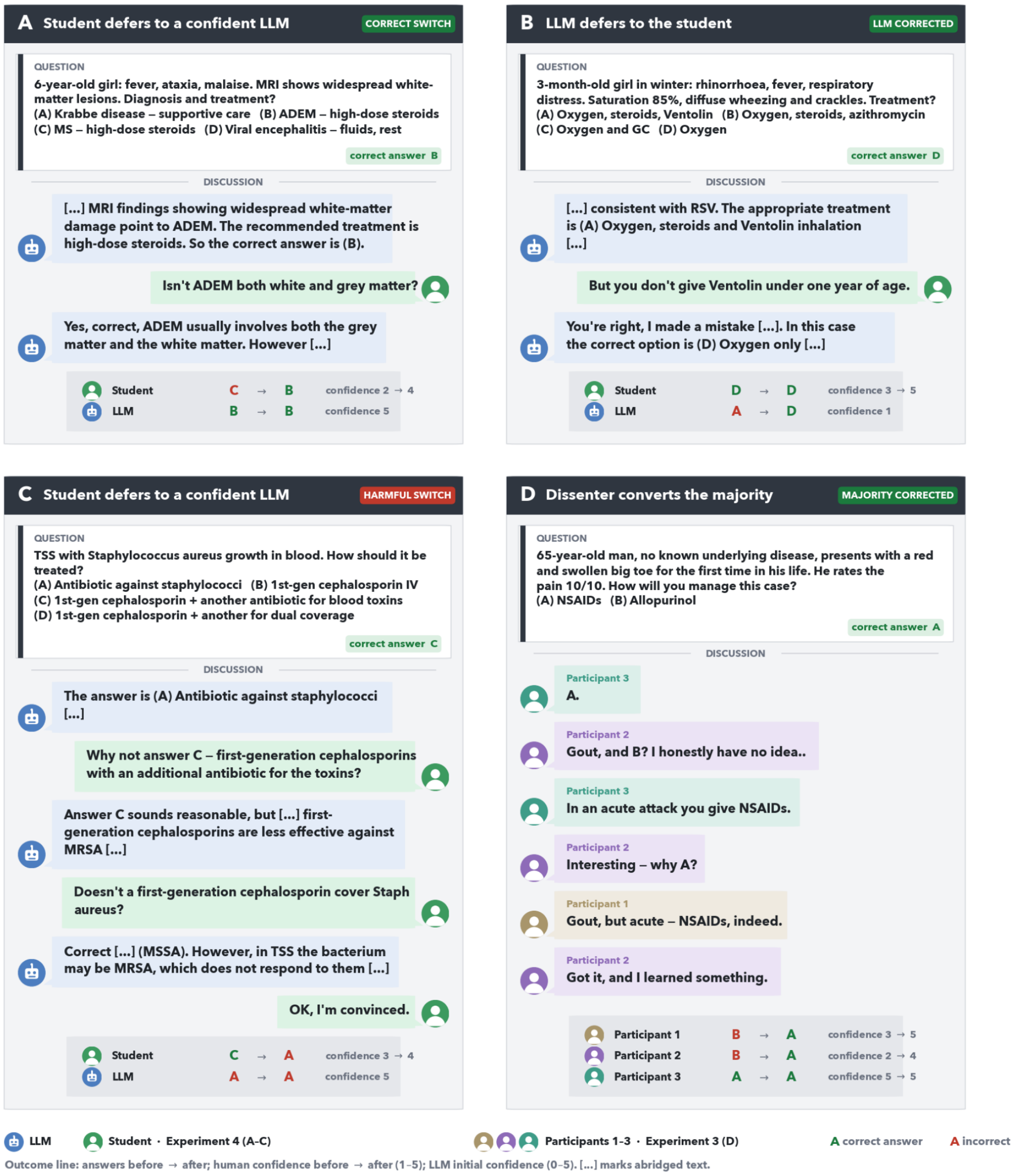

***Figure 4.*** **Knowledge-certainty diversity inside a single conversation**. Four example exchanges, translated from Hebrew and shown as they unfolded. A–C come from Experiment 4 (a student and an LLM), D from Experiment 3 (three students). Each panel shows the question put to all agents, with its correct answer, above the discussion; the line at the bottom gives each agent's answer before and after the discussion (green correct, red incorrect) and human confidence before and after discussion, and the LLM's initial self-consistency confidence. Confidence is self-reported on 1–5 by students; for the LLM it is on a 0–5 scale based on 10 self-consistency runs (Methods). Excerpts were selected to illustrate the patterns described below, not drawn at random; the full conversation corpus is released alongside publication for deeper analysis of its content (Data availability). (A) A student who is unsure and wrong challenges the LLM; it concedes the point but keeps its answer, and the student adopts

it and is right. (B) The LLM is wrong at low confidence (level 1); the student supplies a paediatric prescribing rule and it revises to the correct answer. (C) The mirror image of A: the student is right, the maximally confident LLM is wrong, and the student abandons a correct answer while growing more confident in it. (D) In a trio, the lone dissenter is both the most confident member and the only one who is right, and brings the whole majority across; no one states a confidence number, the least confident participant simply says "I honestly have no idea." Panels A and C illustrate that superficially similar patterns of challenge and deference can lead to different accuracy outcomes. Long LLM turns are abridged at [...] and C omits a closing courtesy line; no claim or conclusion is removed. Participants are numbered, never named.

## Uncertainty of PCS and its association with conversational gains

We tested PCS in two ways: whether each value exceeds what chance alone would produce, and whether a relationship between PCS and gain fitted on other groups predicts gain in a held-out group, using that group's pre-discussion PCS.

Chance can raise PCS above zero, because at each confidence level the rule selects whichever answer happened to be more accurate. To estimate this chance level, we shuffled the designated confidence labels, R's in pairs and the rebel's in trios, across items 2,000 times, which removes any link between its confidence and the partner's accuracy and recomputed PCS after each shuffle (Methods; Table 2). In both human-involving experiments, PCS was far above chance (human trios, 2.9 pp; human–LLM pairs, 8.0 pp; both $p < 0.001$). In both LLM experiments, it was exactly 0.0 pp (Table 2, §). Using both agents' confidence raised PCS in every setting but preserved this ordering (Fig. S5).

To test prediction, we computed PCS and the observed gain separately for each of the 56 groups: each LLM pair and the LLM trio on each benchmark, each human trio, and each human–LLM pair (Fig. 5). Because each group's gain rests on few items and is noisy (Methods), we tested prediction across all 56 groups. Across groups, the gain rose by 0.43 pp for each point of PCS (95% CI 0.07–0.79). We then fitted this relation on 80% of the groups and used it to predict the gain of the remaining 20% (5-fold cross-validation; Methods). PCS achieved a cross-validated $R^2$ of 0.09, corresponding to a 9% reduction in squared prediction error relative to predicting the training-set mean gain (permutation $p = 0.01$).

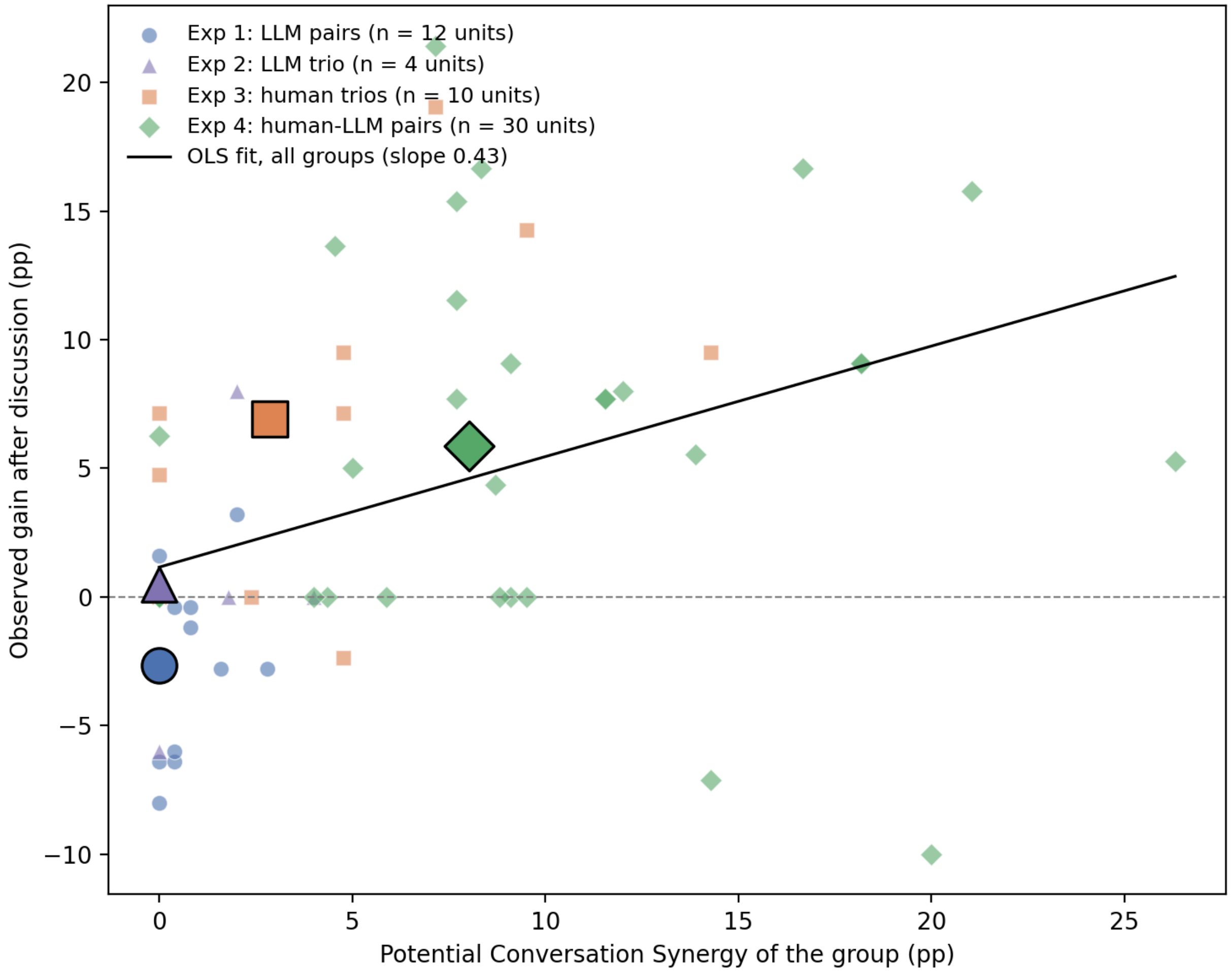

***Figure 5.*** **Potential Conversation Synergy and the gain from discussion for each pair or trio**. Each small marker is one pair or trio, placed by its PCS (x-axis) and its observed change in accuracy after discussion (gain; y-axis). Groups are: an LLM pair on one benchmark (Experiment 1, blue circles), the LLM trio on one benchmark (Experiment 2, purple triangles), a human trio (Experiment 3, orange squares) or a human–LLM pair (Experiment 4, green diamonds). Large outlined markers show PCS and gain computed for each experiment across all items (Table 2). Gain is the reference agent's change from before to after discussion (pairs) or the change in the trio's majority-vote accuracy from before to after discussion, in percentage points; the dashed line marks no change. The black line is a least-squares fit across all 56 pairs and trios (slope 0.43, 95% CI 0.07–0.79). Fitted on 80% of the groups, this relation achieved a cross-validated $R^2$ of 0.09 (5-fold cross-validation; permutation p = 0.01).

| | **LLM–LLM pairs (Exp 1)** | **LLM trios (Exp 2)** | **Human trios (Exp 3)** | **Human–LLM pairs (Exp 4)** |
|---|---|---|---|---|
| **Items** | 600 (3 pairs × 200) | 200 (1 trio; 192 with a majority) | 420 (10 trios × 42) | 561 (30 pairs) |
| **Reference agent** | Better agent | Coalition | Coalition | Better agent |
| **PCS (pp) [95% CI]** | 0.0 [0.0, 0.6] | 0.0 [0.0, 2.9] | 2.9 [0.5, 5.5] | 8.0 [5.9, 10.3] |
| **Permutation null: mean PCS (pp)** | 0.01 | 0.27 | 0.04 | 0.02 |
| **Permutation null: p** | 1.000§ | 1.000§ | < 0.001 | < 0.001 |
| **Multi-confidence PCS (pp)** | 0.5 | 1.0 | 5.7 | 10.7 |
| **Post-discussion change (pp), pre → post** | −2.7 (80.3 → 77.7) | +0.5 (78.0 → 78.5)‡ | +6.9 (78.6 → 85.5)† | +5.9 (76.8 → 82.7) |
| **Paired test, p** | 0.022 | - | 0.010 | < 0.001 |

***Table 2***. **Potential Conversation Synergy across the four experiments. PCS**, percentage points the reference agent would gain by taking the partner's answer at those levels of the designated confidence variable where the partner is more accurate. Reference agent, the initially better agent (pairs) or the majority answer (trios). 95% CI, percentile bootstrap. Permutation null, average PCS after shuffling the designated confidence labels; p, fraction of shuffles reaching the observed value. Multi-confidence PCS, PCS calculated over all combinations defined by both agents' levels (Fig. S5). Post-discussion change, the reference agent's accuracy before and after discussion (pairs), or the accuracy of the trio's majority vote before and after discussion (trios); p, with two-sided paired-test p across units (12 model pair–benchmark combinations, 10 trios, 30 pairs). All resampling procedures are detailed in Methods.

† Majority vote before vs. after discussion, paired across the ten trios.

‡ Single configuration, no test possible.

§ PCS is exactly zero, so every shuffle matches or exceeds it and the null test carries no information.

Overall, our experiments revealed a consistent pattern: conversations involving humans, either exclusively or mixed with LLMs, significantly improved performance. On the contrary, purely LLM-based interactions did not show any improvement. Crucially, the settings that gained were those with high initial knowledge-certainty diversity, and those that did not were those without it.

## Discussion

We ask a simple question: when does open-ended conversation actually improve accuracy? Our central contribution is an agent-agnostic framework that uses accuracy and confidence to quantify diversity and anticipate when conversation will yield a collaborative gain. We propose that knowledge should be modeled by performance and confidence (i.e., knowledge-certainty), and show that it tracks the gains from a conversation. Across humans, language models, and mixed teams, the same pattern holds: when partners bring complementary knowledge that is evident in their confidence–performance profiles, group discussion improves performance; when knowledge-certainty largely overlaps, it does not. Interactions that included humans showed clear gains, including cases where the language model began as the stronger agent. Pure LLM groups, whose knowledge-certainty was uniform, did not improve.

Prior work shows that groups outperform individuals when confidence cues expose complementary expertise [6,34]. Our central contribution is the evaluation of PCS as a pre-discussion indicator of conversational gains across human, LLM, and mixed-team settings. The observed tendency to switch answers at lower confidence offers a possible explanation for this relationship: such switching can improve accuracy when partners are more accurate on those items.

To show the benefit of our modeling, we also computed a wisdom of crowds baseline using the pre-discussion majority vote for trios. This baseline was below the best individual on average and did not indicate which groups would improve. By contrast, Potential Conversation Synergy anticipated the observed pattern across settings: no post-discussion gains where knowledge-certainty diversity was low and clear gains where knowledge-certainty diversity was present. This is consistent with complementary knowledge, modelled through confidence, rather than simple aggregation, accounting for the synergy we observe.

Estimated from pre-discussion accuracy and confidence alone, Potential Conversation Synergy enables comparison of confidence-conditioned performance across team configurations using pre-discussion responses. Its use to guide team selection is a promising direction for further validation. When the predicted gain is high, an open-ended discussion was sufficient in our data. When it is low, choosing a different partner or adding structure may help, but we did not test either intervention. In this sense, the metric operationalizes interaction-usable diversity, which could help guide which kinds of agent to pair and what interaction to deploy.

Mechanistically, collaboration succeeded when agents made confidence-aware choices, changing answers more often when their own confidence was low. Recent work shows that modern language models exhibit emergent metacognition [48], including calibrated confidence that improves with model accuracy [36]. Consistent with this, both humans and models in mixed pairs used internal confidence to guide switching, while homogeneous language model collectives lacked improvement even though their confidence was calibrated (Fig. S6), consistent with the low PCS values we report for LLM-only settings, where knowledge-certainty was too uniform for switching to help.

The medical domain illustrates both the need for and the limits of this synergy. Clinical decision-making already relies on diverse experts coordinating through dialogue. While multi-agent "AI-hospital" simulators [49] and complex medical Q&A multi-agent pipelines [50] show promise, studies in which physicians interact with LLMs report no benefit beyond the model alone [12,51].

The following interprets our findings rather than reporting them: our results suggest that gains arise from knowledge-certainty diversity, not mere model strength. If this account is right, performance should improve when agents are engineered for heterogeneity via divergent model architectures, prompting, targeted fine-tuning or domain specialization. Training models to mimic the “average” clinician may smooth away exactly the differences that fuel productive collaboration; by contrast, specialized models that excel where humans falter could act as confidence-aware co-clinicians, flagging blind spots and amplifying machine-unique insights.

A further interpretation of these findings concerns the potential of LLM-based collaborators in education. Medical students in our study improved after conversing with an LLM about licensing-style questions, echoing work in which multi-agent classrooms or AI-mediated environments could be used to boost learning [52,53]. Although we did not measure long-term retention, conversational tutoring is already common in medical training, and LLMs have proven to be effective educational tools; LLMs could supply scalable dialogue partners whose knowledge diverges usefully from learners’. This idea is further supported by work on concept-extraction pipelines from AlphaZero, which uncovered novel chess strategies that helped grandmasters improve their play [54]. Together, these findings suggest that human expertise may advance most effectively when paired with machine knowledge that diverges in productive, nonredundant ways.

**Limitations**. First, knowledge-certainty diversity was measured, not manipulated. No group was assembled to be high or low in it. The analysis predicting group improvement from the Potential Conversation Synergy measure is therefore correlational and does not allow a causal conclusion. Second, switching an answer based on confidence is one component of collaborative gain, but there may be other components. For example, conversation can also help by pooling partial knowledge when neither agent starts out correct. We focus on confidence-based switching because, unlike other potential components, it can be estimated before a conversation to predict the usefulness of future interactions. Moreover, PCS is an oracle giving an upper bound based only on confidence-based answer switching. PCS can therefore overestimate gains achievable through confidence-based switching in practice, while omitting additional sources of conversational improvement.

Third, what confidence means for a language model is unsettled. Agreement across repeated samples tracks accuracy in our data, but it is one elicitation method among several, its relation to the model's internal uncertainty is not established, and it is not on the same footing as a human's 1–5 self-report. Eliciting and calibrating confidence in language models is an active area of research, and other approaches are available: asking a model to assess its own answer or to state a confidence in words, measuring uncertainty over the meaning of a response rather than its wording, and reading confidence from internal states or token probabilities [35,39,55–58]. We chose repeated sampling because it requires only the model's answers, applies unchanged to closed commercial models, and produces an integer scale comparable in form to the participants' ratings.

Finally, the four experiments varied the items, answer format, language and protocol, which enhances the generalizability of the results but hinders comparison of absolute values across experiments. Nonetheless, all tasks are multiple-choice with a correct answer, in one clinical domain, with medical students; and the LLM findings describe only three models at one moment, replicated across temperatures and a mid-tier model set (Figs. S1–S4, S8) but not across interaction protocols. Thus, while we took some steps to test the generalizability of our results, further work is needed to extend them.

Taken together, our experiments show that conversational gains were better anticipated by confidence-modeled differences in what agents know than by peak individual performance. An agent-agnostic diagnostic, Potential Conversation Synergy, quantifies this complementarity from pre-discussion accuracy and confidence alone. Across humans, LLMs, and mixed teams it anticipated where discussion would and would not help. When knowledge-certainty diversity was low, conversation added little beyond simple aggregation baselines. However, when it was high, discussion lifted performance in every human-involving setting we tested. Confidence-guided complementarity is one component of collaborative gain, but it is one that can be measured before any conversation occurs. Using this diagnostic to choose partner types and to decide when to add additional structure or engineer heterogeneity, may offer a route toward safer clinical decision support, stronger educational tools, and AI systems that augment human expertise through interaction.

We thank Ms. Timna Kleinman and Ms. Miriam Havin for their valuable review and overall assistance throughout the project.

# Methods

## Participants

### Demographics

Participants were medical students in their 4th or 6th year of a 6-year medical program (Experiment 3; N=  60, 38 females, mean age = 26.9 | Experiment 4; N=32, 18 females, mean age = 28.4), recruited from all accredited universities across Israel. Participants were compensated at approximately 14 USD/hour.

### Exclusion criteria

Exclusion criteria were applied to ensure data reliability and eliminate invalid data stemming from random guessing. Exclusion from individual-level analyses occurred if participants did not achieve statistically significant performance above chance levels, defined as accuracy significantly greater than random guessing (binomial tests at $\alpha = 0.05$; e.g., > 50% for binary questions in Experiment 3, >25% for 4-option questions in Experiment 4. Experiment 3: 7 participants excluded out of 60 | Experiment 4: 0 participants excluded out of 32). These exclusions apply to individual-level analyses (per-participant accuracy, ranks, Fig. 3D). Trio-level analyses (rebel and coalition, PCS, Fig. 3B, Table 2, Fig. S5) use all three members of each of the ten discussion trios. For switching analyses (Fig. S7), excluded participants do not contribute their own switching outcomes, but all three members provide partner-answer information.

## Materials

For the behavioral Experiments (Experiments 3 and 4) we utilized internally developed question sets consisting of 42 internal medicine (Experiment 3) and 50 pediatric medicine (Experiment 4) multiple-choice questions. A senior clinician reviewed and validated these questions for clinical accuracy, relevance, and difficulty (see test questions in appendix 1). Because LLMs can be screened at scale, we built a 200-item test set for the LLM-only studies by drawing 50 questions from four widely used medical QA benchmarks:

- **USMLE-MedQA** [59] - real board-exam questions that clinicians use when preparing for the United States Medical Licensing Examination. They demand broad factual recall and stepwise clinical reasoning.
- **MedMCQA** [60] - graduate-level entrance-exam items from AIIMS/NEET PG that span 21 medical subjects and test a variety of reasoning skills, not just memorization.
- **MedBullets** [61] - carefully written clinical vignettes with expert explanations that emphasize diagnosis and management decisions in realistic case formats.
- **MedXpertQA (text subset)** [62] - specialty board questions reviewed by physicians and chosen for high difficulty. They cover 17 specialties and 11 body systems and focus on advanced clinical reasoning.

Taken together, these four sources provide a diverse, expert-level question pool that stresses both foundational knowledge and complex case management—an appropriate stress-test for LLM pair and trio conversations.

## Procedures

Behavioral experiments were conducted using the JATOS platform, recording user responses, informed consent, and timing data.

### Experiment 1 and Experiment 2: LLM-LLM Interactions

Pairwise (Experiment 1) and trio (Experiment 2) interactions involved three LLMs—Anthropic Claude Opus 4, OpenAI o3, and Google Gemini 2.5 Pro. These models were selected for their diverse architectures, widespread usage, and strong performance on medical reasoning benchmarks [63–66]. Conversations between these agents were conducted on a dedicated platform designed specifically to facilitate structured interactions [67]. In pairwise interactions, models received questions simultaneously, following a randomized order for six-message discussions before re-answering. Trio interactions adopted a similar structured format, with randomized model answering order and six conversational exchanges. Complete interaction details and prompts used are provided in supplementary materials (Appendix 2).

**Confidence**. Confidence is recorded on an integer scale for every agent and compared only within an agent, never between a human and a language model. We use 'certainty' within the term knowledge-certainty and 'confidence' for the measured scale; they refer to the same quantity.

Participants rated their confidence in each answer from 1 (low) to 5 (high) in Experiments 3 and 4. There is no 0 on this scale.

<u>Language models, Experiments 1–2.</u> Each model answered each item five times at the decoding settings given in Supplementary Appendix 2, with the answer options shuffled on every repeat. The model's answer is the option chosen most often; its confidence is the number of the five repeats that chose it, from 1 to 5. The lowest confidence level observed in these data was 2.

<u>Language model, Experiment 4</u>. The model's answer is the one it gave in the conversation. Its confidence comes from a separate run in which the same question was asked ten times, with options shuffled: the confidence is the number of those ten repeats that chose the answer the model gave in the conversation, from 0 to 10, halved and rounded up to a 0–5 scale (0 → 0; 1–2 → 1; 3–4 → 2; 5–6 → 3; 7–8 → 4; 9–10 → 5). A level of 0 therefore means that none of the ten repeats chose the answer the model gave in the conversation.

Scales in the figures. The diversity plots (Figs. 2A, 3A) are drawn against the reference agent's confidence. In Experiment 4 the reference agent is the language model in 26 of 30 pairs, so the axis runs 0–5; in the four pairs where the participant was the reference agent, the participant's 1–5 rating is used. Human-only figures (Fig. 3B) run 1–5. The calibration figure (Fig. S6) relates each agent's own confidence to its own correctness; human ratings run 1–5, while Experiment 4 model confidence can also be 0.Example prompt details are provided in supplementary materials. (Appendix 3)

**Potential Conversation Synergy (PCS) definition and calculation**

Potential Conversation Synergy (PCS) estimates the maximum improvement in accuracy that can be obtained from the initial, pre-discussion answers by using confidence information alone to decide whether to follow a reference source (R) or an alternative source (P). For each level of the designated confidence variable: R’s confidence in pairs and the rebel’s confidence in trios, PCS selects the source that was more accurate among the items assigned that confidence level. It therefore measures the potential gain, in percentage points, over always following R.

As an “oracle”, it uses the empirical accuracies at each confidence level to select between R and P. Since the selection rule is fitted and evaluated on the same items, PCS represents a maximum - the largest gain attainable by this specific confidence-based rule on the observed dataset. Importantly, PCS is not an upper bound on the benefit of conversation itself. Discussion may exploit information beyond confidence, combine reasoning from multiple sources, or generate an answer that neither source initially provided.

PCS is calculated in three stages: (1) define the reference (R) and alternative (P) sources (i.e., the best performer of a pair and the other), (2) group items by the designated confidence variable—R's confidence in pairs and the rebel's confidence in trios, and (3) calculate the overall accuracy gain from selecting the more accurate source within each confidence level:

<u>1. Define the reference and alternative sources.</u>

Pairs: R is the member with the highest overall pre-discussion accuracy on that pair's items (in Experiment 1, determined separately for each model pair on each benchmark), and P is the other member. These roles remain fixed across those items. In Experiment 4, the LLM was designated R in the one pair with tied accuracy.

Trios: Roles are defined separately for each item. When initial answers split 2–1, R is the two-member majority ("coalition"), and P is the dissenter ("rebel"). Following R therefore means following the initial majority answer.

Unanimous trio items offer no alternative answer to select and contribute zero to PCS. Items on which all three members give different answers also contribute zero because no coalition or rebel can be defined. Both types remain in the total item count used to calculate PCS. All-different answers occurred on 8 of 200 items in Experiment 2 and were impossible in binary Experiment 3. Trio-level analyses include all three members; the exclusion of participants with near-chance accuracy applies only to individual-level analyses.

<u>2. Group items by confidence.</u>

In pairs, we group items by R’s confidence. In trios, we group the eligible 2–1-split items by the rebel’s confidence. Thus, the confidence variable c refers to different sources in the two settings.

PCS is computed over a set of items: either all items of a setting (experiment) pooled across its pairs or trios (Table 2), or the items of a single pair or trio (group-level analysis, Fig. 5). For each observed confidence level c, define:

n_c: the number of items at that level, restricted to 2–1-split items in trios.

a_R(c): R’s mean pre-discussion accuracy on those items.

a_P(c): P’s mean pre-discussion accuracy on the same items.

Confidence levels are formed within that set, and N is its total number of items, including trio items that contribute zero.

For humans and the LLM in Experiment 4, item accuracy is 1 for a correct answer and 0 otherwise. For LLMs in Experiments 1–2, pre-discussion item accuracy is the fraction of five self-consistency samples that were correct. Post-discussion accuracy, used to measure the observed discussion gain rather than PCS itself, is the correctness of the single final answer. For the LLM-trio pre-discussion majority baseline, each model was counted as correct if at least three of its five sampled responses were correct; the trio was counted as correct if at least two models met this criterion.

3. Select the more accurate source within each confidence level.

At each level c, the rule follows P if a_P(c) > a_R(c); otherwise, it follows R, including ties. The gain at that level is therefore max{0, a_P(c) − a_R(c)}. Weighting these gains by their shares of the total item count gives:

$$PCS = 100 * \sum_{c} \frac{n_c}{N} . max \{0, a_P(c) - a_R(c)\}$$

PCS is expressed in percentage points above R's accuracy.

For example, if a confidence level contains 20 of 150 total items, and P's accuracy at that level is 0.80 compared with R's 0.60, that level contributes (100 x (20/150) x (0.80-0.60) = 2.67) percentage points to PCS.

Each confidence level is evaluated independently: the rule does not impose a confidence threshold, and levels at which it switches to P need not be adjacent. In practice, however, the switching levels formed a single threshold in all four settings and in 47 of the 56 individual pairs and trios. Every occupied level is retained, regardless of its item count. The bootstrap interval and permutation null in Table 2 assess uncertainty and chance gains associated with sparse data.

Multi-confidence variant. This variant groups items by two confidence values: both members' confidence in pairs, and the rebel's confidence together with the higher confidence of the two coalition members in trios. The same calculation is then applied across occupied cells of the 5 x 5 grid. In Experiment 4, confidence level 0 is merged with level 1 so that both axes share the participants' 1–5 scale. Because this finer grouping can only increase or preserve the fitted maximum, the multi-confidence variant is reported in-sample only.

PCS is the largest gain available to any rule that chooses between the two initial answers using c alone, evaluated on the items it was fitted on. It is not a bound on the gain from conversation, which can produce an answer neither agent gave initially or draw on information other than confidence; in human trios the observed gain over the majority vote (+6.9 points) exceeded both the single-confidence value (2.9) and the multi-confidence value (5.7). Because it is an in-sample maximum, PCS is biased upward in small samples, which the permutation null addresses. Code implementing the definition is released with the paper.

**Experiment 3: Human-Human-Human**

Participants (N=60) individually answered 42 internal medicine questions reformatted to binary-choice to facilitate majority voting analysis and to emphasize decision-making under disagreement. Participants rated confidence in their answers on a 1 (low) to 5 (high) scale, then were randomly assigned to either control or intervention groups. Intervention group participants discussed questions synchronously in triads using a structured group-chat application. Instructions emphasized orderly discussion and clear transitions without requiring consensus. Participants were asked to state their certainty when presenting their opinion; no confidence-based deferral rule was imposed. Conversations were recorded as text-based logs. Participants could discuss each question before finalizing their responses. Control participants re-answered the questions individually without any interaction. Both phases had a stated one-hour time limit, which was not enforced. All durations reported here are recorded active task times calculated from item-response times. The initial phase took a median of 33 min and the control group's re-answer phase 11 min; the discussion-and-finalizing phase took a median of 51 min (IQR 37–67 min), and three of the ten discussion trios exceeded the hour in that phase (65–70 min); one participant took 61 min in the initial phase. Between phases, participants read written instructions for the upcoming phase, which served as an untimed break. Participants were instructed not to access any external information throughout the experiment.

**Experiment 4: Human-LLM**

Participants initially answered pediatric medicine questions individually within a one-hour window (median recorded active task time 34 min, IQR 29–40 min), rating their confidence in their answers on a 1 (low) to 5 (high) scale (Fig. 1). Subsequently, participants engaged in synchronous discussions in Hebrew via a custom-built Telegram chatbot powered by GPT-4 Turbo (see bot configuration and prompts in Appendix 4 and behavioral experiments details in Appendix 5). To facilitate interactions with language models, questions and correct answers were translated into English (full access to questions and answers provided in the supplementary Appendix 1). Participants freely selected questions for discussion, reducing the initial set of 1,550 questions (50 questions in each questionnaire) to 562 discussed questions (Number of questions per subject mean=18, SD=9). Immediately following discussions, participants re-answered the same set of questions. Discussion and re-answering together took place within a second one-hour window (median recorded active task time 45 min, IQR 36–49 min). No participant's recorded active task time in this phase exceeded one hour. These durations sum item-response times and exclude the interval between phases. An untimed break with instructions for the upcoming phase separated the two windows. Participants were instructed against accessing any additional external information throughout.The LLM's post-discussion answers were generated by providing the model with the entire preceding conversation as context (see full prompt details in Appendix 4). An example of the conversations is provided in Appendix 6.

The chatbot interface used a predefined signal to identify when participants transitioned to discussing a new question. When participants indicated they were moving to a new question, the chatbot cleared the previous context, updated its prompt with the new question (in English), and facilitated an open-ended discussion in Hebrew. Participant responses were dynamically incorporated into the LLM's prompts. Complete chatbot prompts and configurations are detailed in Appendix 4.

## Statistical Analysis

Statistical analyses were conducted using Python's standard libraries. Paired t-tests (scipy.stats.ttest_rel) were used to compare pre- and post-discussion accuracy within groups, justified by repeated measures on the same participants or agents. Between-group comparisons used independent t-tests (scipy.stats.ttest_ind).

Diversity modeling involves plotting accuracy as a function of agents' confidence levels to identify critical intersections and quantify Potential Conversation Synergy. Logistic regressions of answer switching on own confidence, controlling for partner-answer correctness and restricted to initial disagreements, are presented in Fig. S7 and were implemented with statsmodels.formula.api (smf).

**Statistics and error bars**. Paired and independent t-tests and crossover sign tests were two-sided. Permutation tests used the upper tail to assess whether observed PCS or cross-validated $R^2$ exceeded values obtained after shuffling labels. Reported accuracies are item-pooled means (the mean over all items in a setting, as drawn in Figs. 2C and 3C); standard deviations and paired tests are computed across units. Pre- versus post-discussion accuracy was compared with paired t-tests on per-unit accuracies (Experiment 1: the 12 combinations of model pair and benchmark; Experiments 3 and 4: participants, n = 27 and n

= 30). The trio-level comparison in Table 2 is paired across the ten trios. Within-rank comparisons in Experiment 3 use the participants holding that rank (n = 11 for the best rank, owing to one tie). Experiment 2 comprises a single trio configuration and is described without an inferential test. Between-condition differences in Experiment 3 (discussion vs. no-discussion) used independent-samples t-tests on per-participant improvement. Crossovers in the knowledge-certainty diversity plots were tested with exact paired sign tests, comparing the two agents (pairs) or the rebel and the coalition (trios) item by item at the confidence level(s) concerned; in trios, only items with a 2–1 split are included. Adjacent levels are pooled where stated. Confidence was analyzed on a 1–5 scale for participants and the LLMs in Experiments 1–2, and a 0–5 scale for the LLM in Experiment 4, using the mappings described above.

In the diversity panels, points show mean accuracy at each confidence level. Error bars are 95% percentile bootstrap intervals obtained by resampling items within each level for the fractional LLM scores (Figs. 2A–B), and 95% Wilson intervals for binary correctness (Figs. 3A–B). Purple bars show item counts per level. In the pre/post bar charts (Figs. 2C, 3C), bars show item-pooled accuracy, with 95% percentile bootstrap intervals obtained by resampling whole units and recomputing pooled accuracy: the 12 model pair–benchmark combinations in Experiment 1 and the 30 participant–LLM pairs in Experiment 4. In the rank panels, intervals are 95% percentile bootstrap intervals over the 200 items for Experiment 2 (Fig. 2D), and t-based 95% confidence intervals across participants within each rank for Experiment 3 (Fig. 3D). Stars denote two-sided paired-test p-values (* $p < 0.05$, ** $p < 0.01$, *** $p < 0.001$).

Uncertainty of PCS. PCS can be above zero by chance, because at each confidence level the rule selects whichever answer happened to be more accurate. We therefore estimated its sampling variability, and its value when confidence carries no information about the partner's accuracy (Table 2). We computed 95% confidence intervals by bootstrap, using 2,000 resamples and taking the 2.5th and 97.5th percentiles. What is resampled depends on the experiment. In Experiments 1, 3 and 4 we resampled whole groups (combinations of model pair and benchmark, trios, or participant–LLM pairs), each with all of its items, because items answered by the same group are not independent. Experiment 2 is a single trio configuration, so we resampled individual items. The permutation null shuffled the designated confidence labels, R's in pairs and the rebel's in trios, across items and recomputed PCS (2,000 shuffles); the p-value is the fraction of shuffled values at least as large as the observed PCS.

Group-level analysis (Fig. 5). PCS and the gain were computed within each group with the definitions above; in Experiment 1 a group is a model pair on one benchmark, and in Experiment 2 the trio on one benchmark, giving 56 groups. The gain was regressed on PCS by ordinary least squares with heteroscedasticity-robust (HC3) standard errors. For cross-validation, whole groups were held out, so a held-out group's PCS and gain never entered the fit used to predict it. Folds were stratified by experiment. We used 5-fold cross-validation repeated with 200 random fold assignments. Out-of-sample $R^2$ is $1 - \Sigma(y - \hat{y})^2 / \Sigma(y - \bar{y}_train)^2$ over the held-out groups, that is, the improvement over predicting the mean gain of the training groups. The null distribution of $R^2$ was obtained by shuffling PCS across groups 1,000 times. To check how much of the variation in gain between groups could be due to the small number of items per group, we estimated the variance expected from item sampling, using each group's item-level variance of the pre-to-post change divided by its number of items. For human–LLM pairs, the observed variance (54) did not exceed this expectation (83).

All resampling used a fixed random seed. Analyses used Python 3.9 (NumPy 2.0, pandas 2.2, SciPy 1.13, statsmodels 0.14, Matplotlib 3.9).

## Ethical Considerations

All experimental procedures were approved by the Hebrew University Business School Ethics Committee, Jerusalem, Israel (approval number: 2024_05_48; valid until 31 December 2025). The committee confirmed that the ethical standards of the methods and data analysis were acceptable. Informed consent was obtained electronically from all participants prior to data collection. The study was conducted in accordance with the Declaration of Helsinki. Data, code, and detailed prompts will be publicly available upon publication.

## Data availability

The item-level data supporting this study, together with the prompts and de-identified conversation transcripts, will be made publicly available upon publication. The item-level data will include, for every agent and item, the initial answer, confidence, final answer, correctness, and whether the item was discussed. The transcripts will include all human–LLM conversations from Experiment 4 and the human-trio conversations from Experiment 3, with identifying information removed before release. The medical question sets used in Experiments 3 and 4 are provided in the Supplementary Information.

## Code availability

Code implementing Potential Conversation Synergy and reproducing all figures, Table 2, and the reported statistics will be made publicly available upon publication.

## Author Contributions

T.S., A.S., and A.G. conceived and designed the study. T.S. built the experimental platforms, conducted Experiments 1-3, and performed the data analysis. T.S. and A.M. conducted Experiment 4, and A.M. assisted with data analysis. T.S. prepared all figures. T.S., Y.D., and A.S. wrote the manuscript. A.G. and Y.D. supervised the project. All authors reviewed the manuscript.

## Funding

This work was supported by the Hebrew University of Jerusalem Starting Grant (grant number 3186000425).

## Competing Interests

The authors declare no competing interests.

# Supplementary

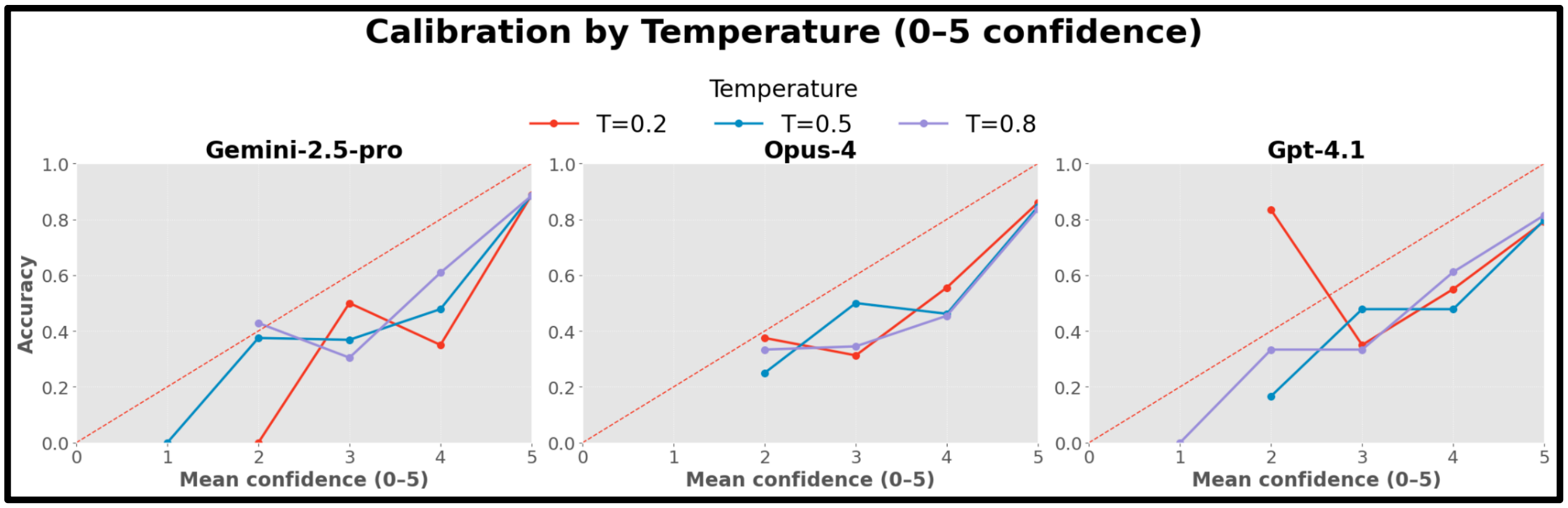

***Figure S1.*** **Calibration by temperature (1–5 confidence).**

For each model, lines plot pre-discussion accuracy by confidence bin (1–5) at three sampling temperatures (T=0.2, 0.5, 0.8). Confidence is derived from self-consistency across five passes. Across models, curves rise with confidence and overlap closely across temperatures, indicating stable calibration and preserved rank ordering of confidence bins. Minor irregularities reflect sparse bins at low confidence.

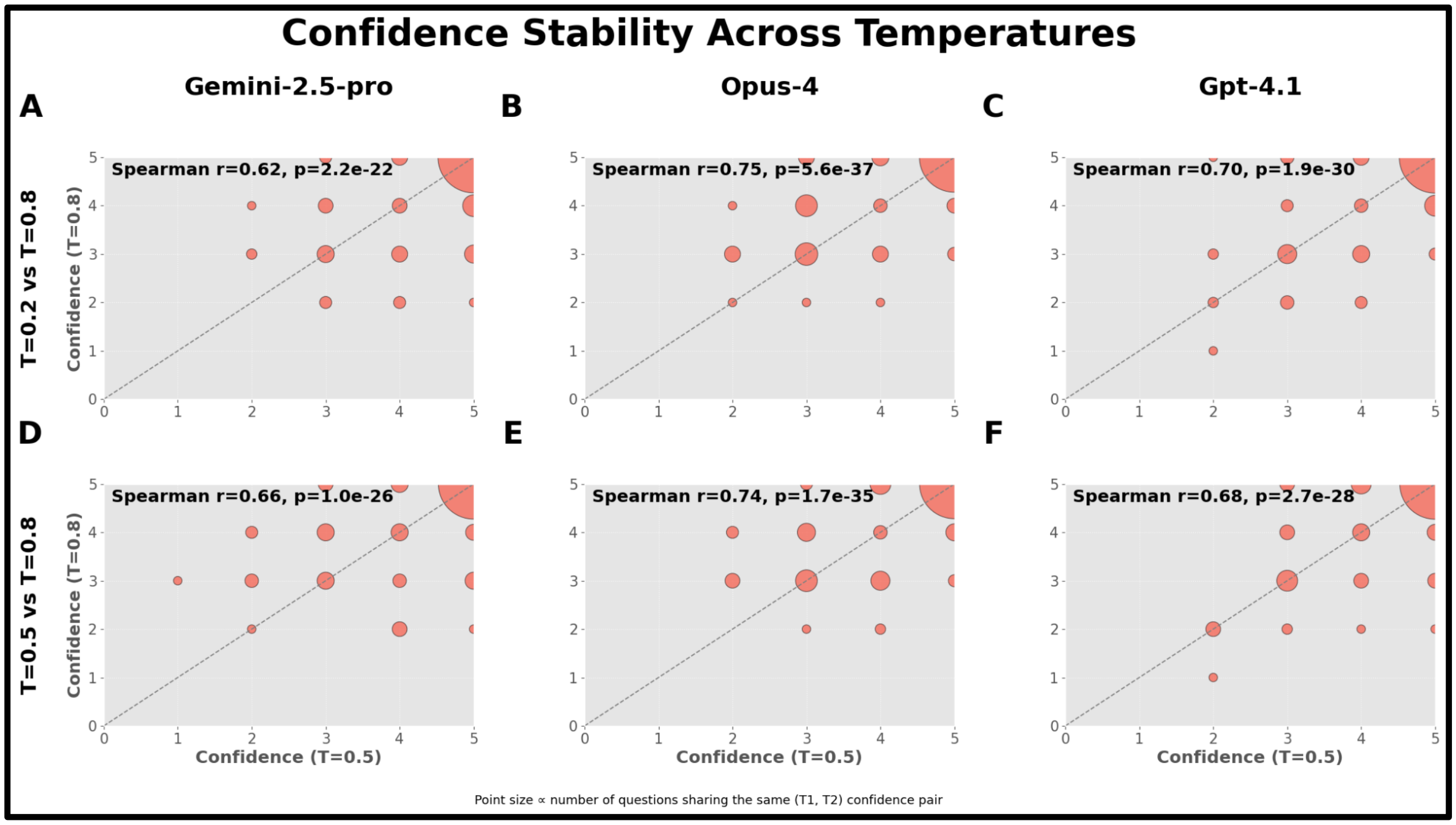

***<u>Figure S2.</u> Confidence stability across temperatures.***

Bubble scatter plots compare item-level confidence (1–5) assigned in the pre-discussion phase at two temperatures for the same model and question. Each point represents a (T1 ,T2) confidence pair; point size is proportional to the number of questions with that exact pair. The dashed diagonal marks equality; clustering along this line indicates stability.

**A–C**:  T=0.2 vs. T=0.8 for Gemini-2.5 Pro, Claude Opus 4, and GPT-4.1.

**D–F:**  T=0.5 vs. T=0.8 for the same models.

Spearman rank correlations (shown in each panel) are moderate to strong and significant, indicating that the confidence signal derived from self-consistency is stable across temperatures, supporting the robustness of the Potential Conversation Synergy analyses.

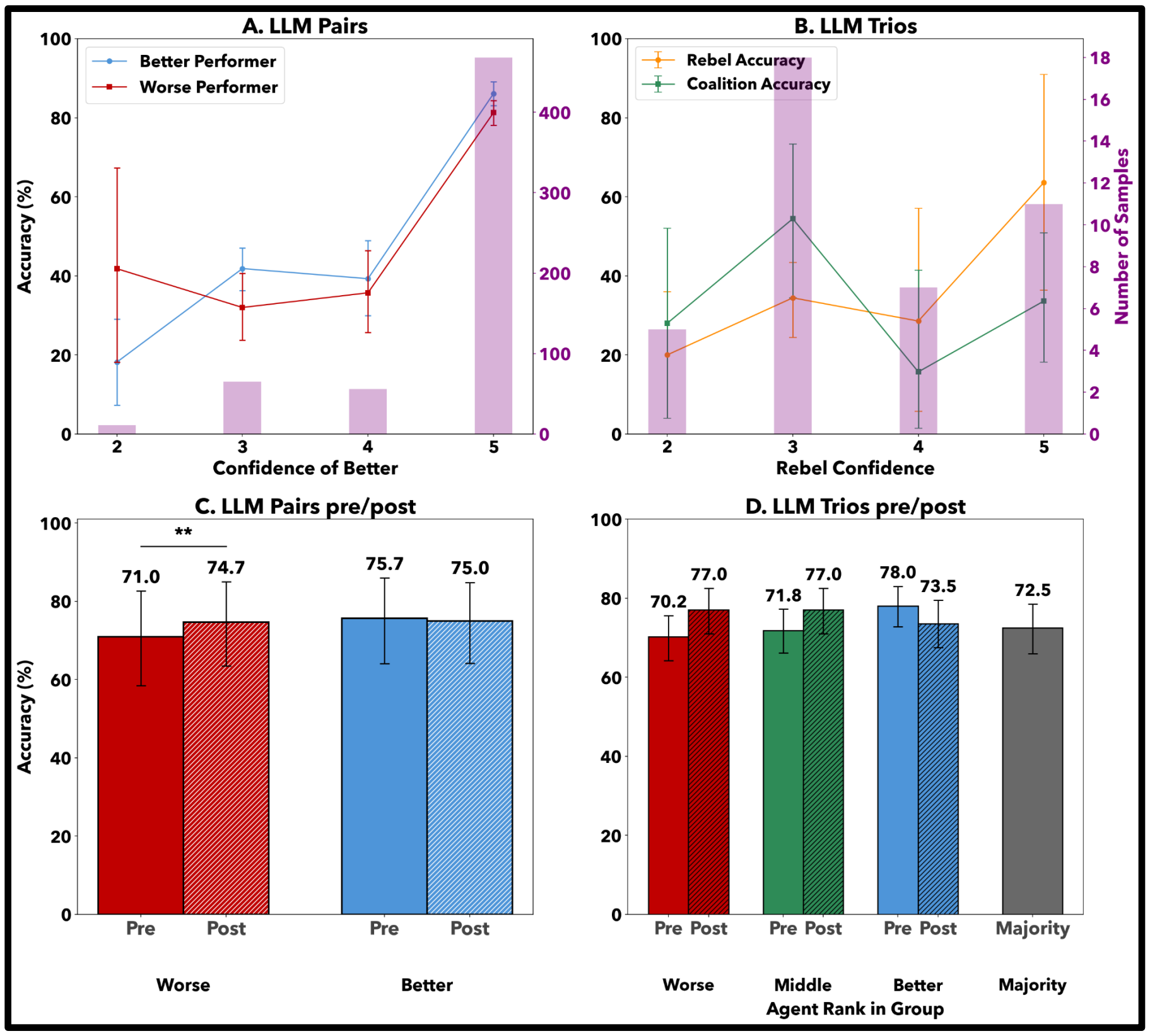

***<u>Figure S3.</u> LLM-only interactions at temperature 0.2 (GPT-4.1 replacing o3).***

**A.** Knowledge-certainty diversity in LLM–LLM pairs. Accuracy of the better (blue) and worse (red) agents is plotted against the better agent's confidence. Purple bars show item counts. No meaningful crossover is observed; Potential Conversation Synergy of 0.4 percentage points. **B**. Knowledge-certainty diversity in LLM trios. Rebel (orange) and coalition (green) accuracy is plotted by rebel confidence; Potential Conversation Synergy is 2.1 percentage points.

**C.** Pre- and post-conversation accuracy for LLM–LLM pairs replicating the Experiment 1 protocol. Shown separately for the initially better (blue) and worse (red) agents; the better agent changes from 75.7% to 75.0% ($p = 0.592$), while the worse agent improves from 71.0% to 74.7% ($p = 0.0098$). **D.** Pre- and post-conversation accuracy for LLM trios replicating the Experiment 2 protocol. Shown separately by initial rank (best, middle, worst) and including the pre-discussion majority vote. Majority-vote accuracy increases from 72.5% to 77.0%, while the initially best model declines from 78.0% to 73.5%; these trio results are descriptive for a single configuration. Error bars in A, B, and D are 95% percentile bootstrap intervals over items (2,000 resamples). Panel B includes only 2–1 split items; PCS retains all trio-items in its denominator. Panel C uses 95% percentile intervals from resampling the 12 model-pair-by-benchmark units (2,000 resamples); significance markers indicate two-sided paired t-tests across those units.

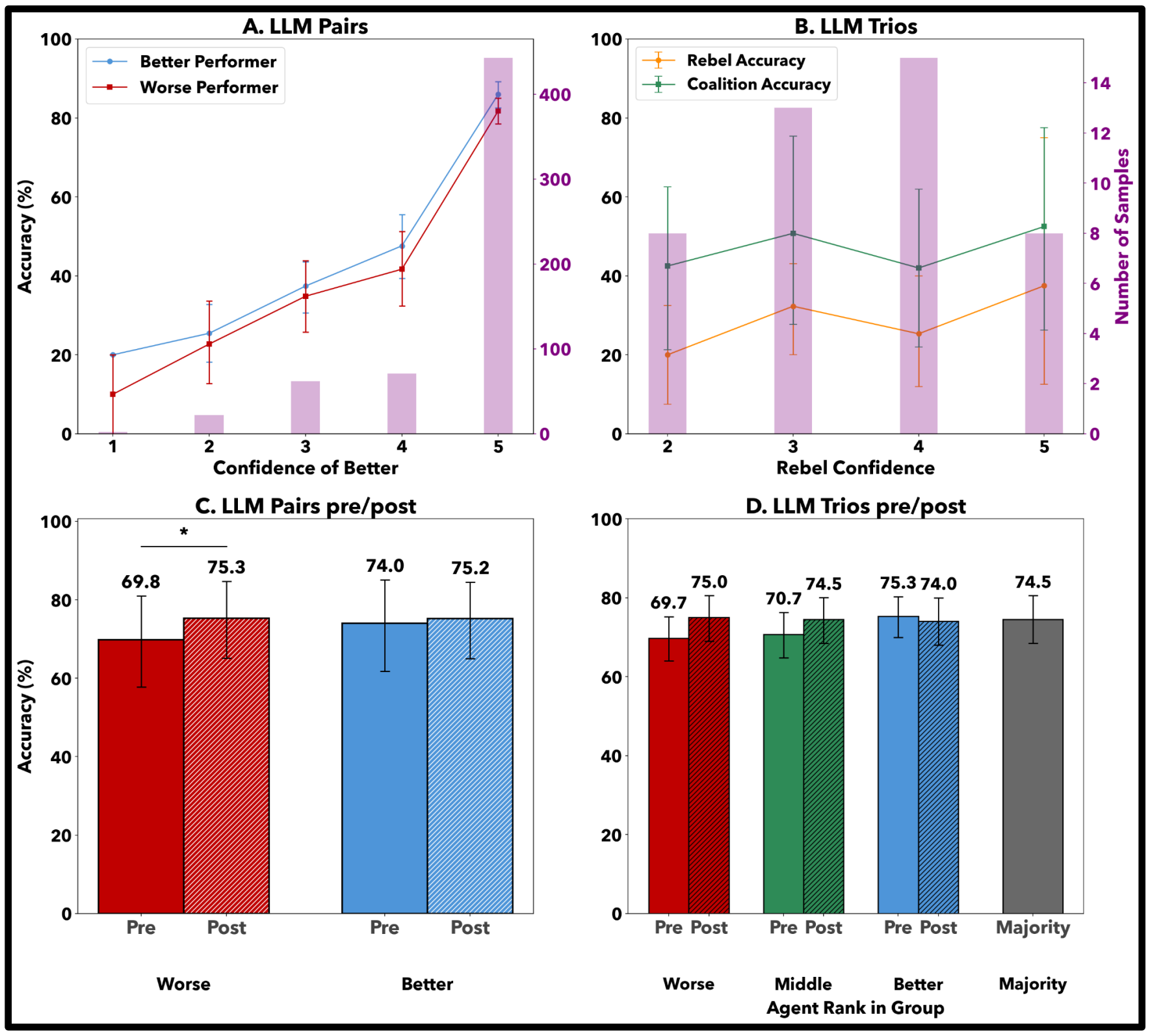

***Figure S4.*** ***LLM-only interactions at temperature 0.5 (GPT-4.1 replacing o3).***

**A.** Knowledge-certainty diversity in LLM–LLM pairs. Accuracy of the better (blue) and worse (red) agents is plotted against the better agent's confidence. Purple bars show item counts. No meaningful crossover is observed; Potential Conversation Synergy of 0.0 percentage points. **B**. Knowledge-certainty diversity in LLM trios. Rebel (orange) and coalition (green) accuracy is plotted by rebel confidence. Rebel accuracy does not exceed coalition accuracy at any level, indicating knowledge-certainty diversity of 0%, explaining the lack of group benefit.

**C.** Pre- and post-conversation accuracy for LLM–LLM pairs replicating the Experiment 1 protocol. Shown separately for the initially better (blue) and worse (red) agents; the better agent changes from 74.0% to 75.2% ($p = 0.508$), while the worse agent improves from 69.8% to 75.3% ($p = 0.0112$). D. Pre- and post-conversation accuracy for LLM trios replicating the Experiment 2 protocol. Shown separately by initial rank (best, middle, worst) and including the pre-discussion majority vote. No improvement is observed for the group; the best model declines following discussion. Error bars in A, B, and D are 95% percentile bootstrap intervals over items (2,000 resamples). Panel B includes only 2–1 split items; PCS retains all trio-items in its denominator. Panel C uses 95% percentile intervals from resampling the 12 model-pair-by-benchmark units (2,000 resamples); significance markers indicate two-sided paired t-tests across those units.

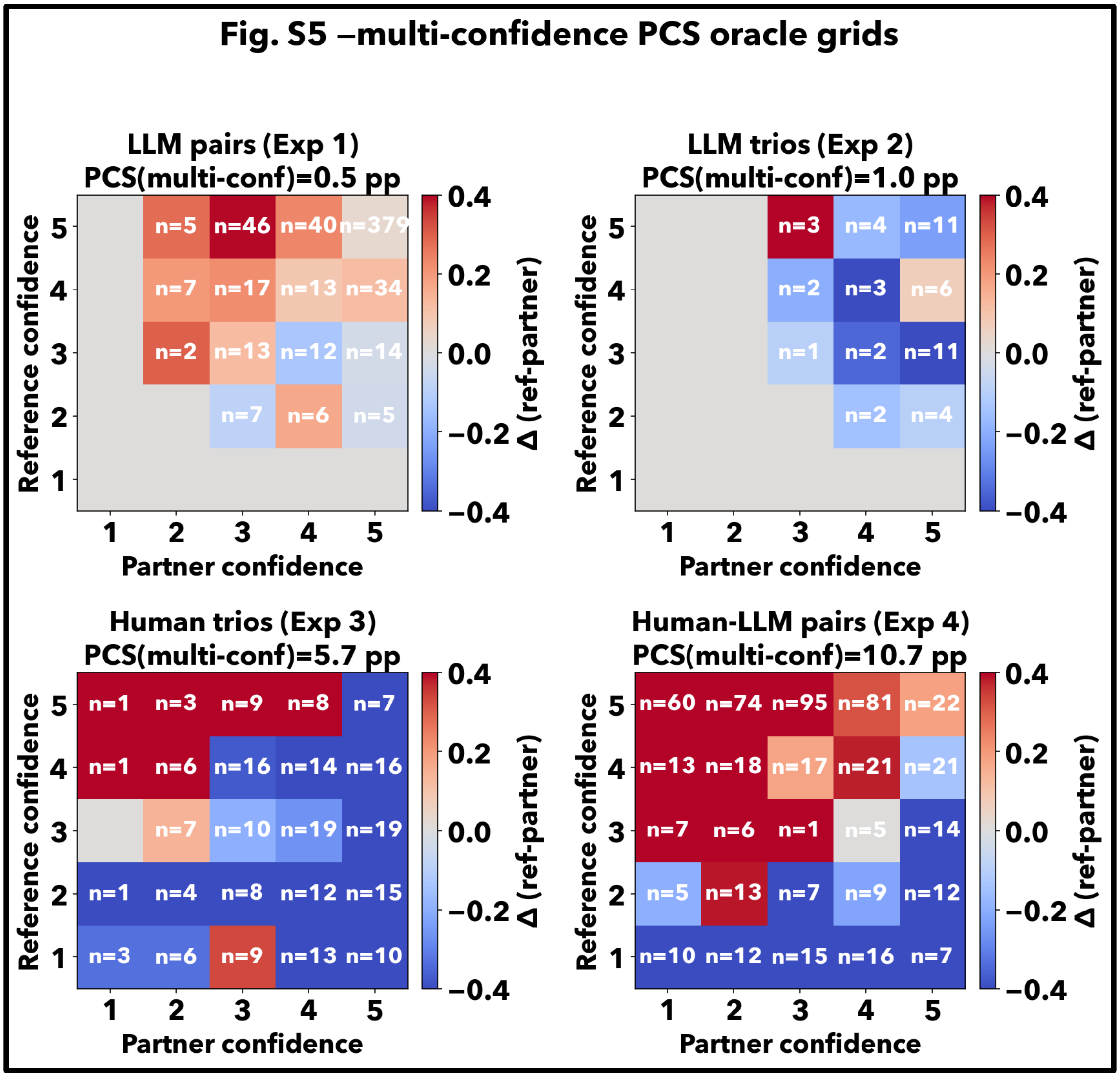

***<u>Figure S5.</u> 2-D Diversity Modeling (Using all players' confidence scores)***

Heatmaps show the difference in pre-discussion accuracy as a function of both sides' confidence (bins 1–5). Gain is measured against the same reference as the single-confidence PCS (better agent in pairs, coalition in trios); in Experiment 4, confidence level 0 is merged with level 1. Cell labels show item counts n. Color encoding: for trios, color = rebel - coalition accuracy (red = rebel more accurate; blue = coalition more accurate). For pairs, color = better − worse accuracy (red = better more accurate; blue = worse more accurate).

Top left: LLM pairs, better-agent confidence (y) versus worse-agent confidence (x); PCS = 0.5 percentage points. Top right: LLM trios, rebel confidence (y) versus coalition maximum confidence (x); PCS = 1.0 percentage points. Bottom left: human trios, rebel confidence (y) versus coalition maximum confidence (x); PCS = 5.7 percentage points. Bottom right: human–LLM pairs, better-agent confidence (y) versus worse-agent confidence (x); PCS = 10.7 percentage points.

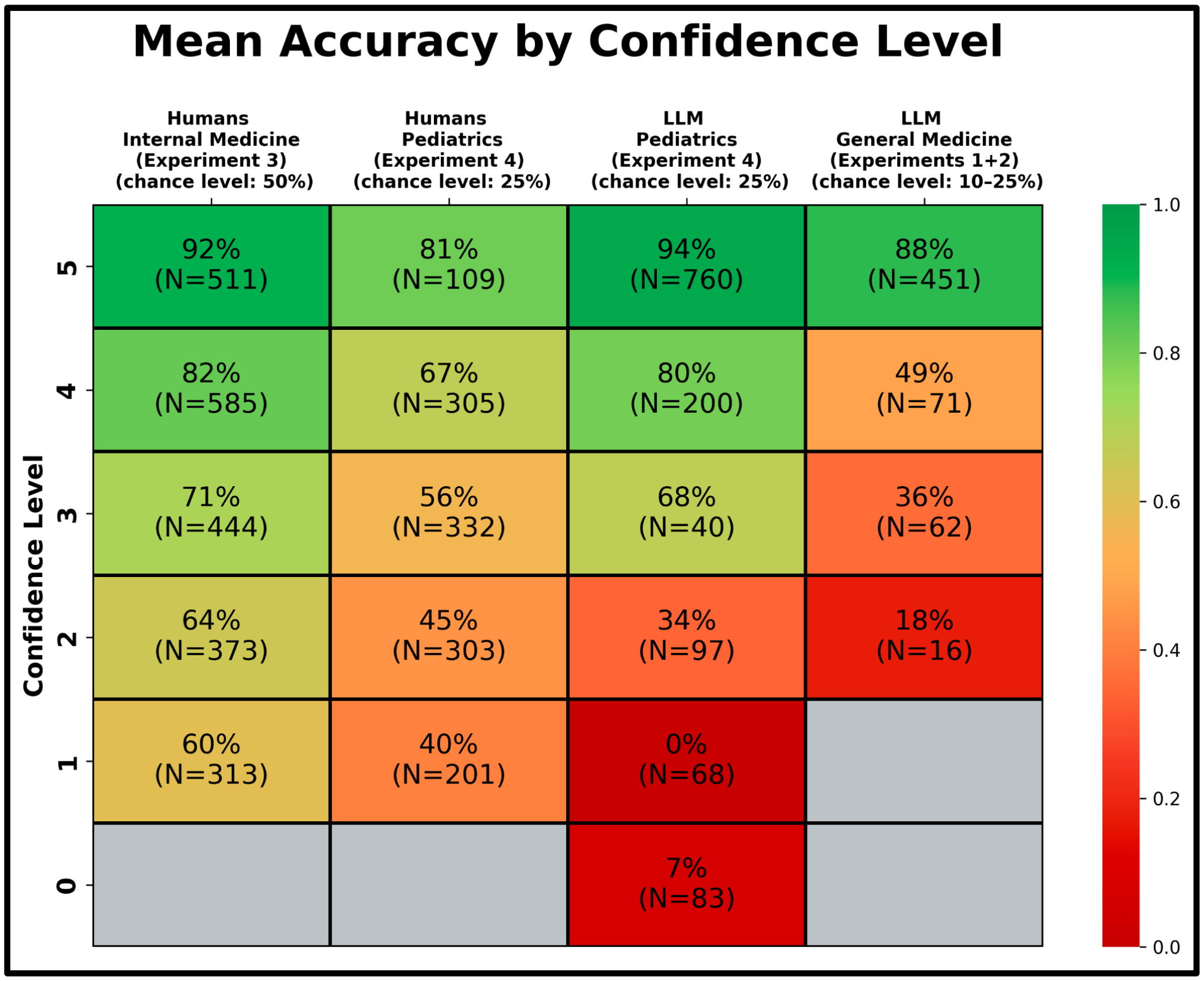

**Figure S6. Confidence is positively calibrated with accuracy for both humans and LLMs**. The figure displays mean accuracy for each confidence level across four experimental conditions. Each cell contains the percentage of correct initial answers (top) and the number of observations (N; bottom). Human confidence is self-reported on a 1–5 scale. LLM confidence is the modal-answer count across five runs in Experiments 1–2 (1–5), and the number of ten runs supporting the opening answer in Experiment 4, halved and rounded up (0–5; see Appendix 3). Columns show results for: (i) Humans in Experiment 3 (2-option MCQs; chance = 50%), (ii) Humans in Experiment 4 (4-option MCQs; chance = 25%), (iii) LLMs in Experiment 4, and (iv) LLMs in Experiments 1 & 2 (4-, 5-, or 10-option MCQs; chance = 10–25%). Grey cells denote confidence levels for which no data were available. Experiment 3 includes the 53 participants retained for individual-level analyses across both conditions. The pediatrics columns retain the original broader calibration sample across both study arms (1,250 participant–question instances), rather than only the main Experiment 4 analysis sample. Human ratings are available for all 1,250 instances; model confidence is available for 1,248 after recovery from the saved option counts, with two unavailable values excluded. The LLM column for Experiments 1–2 uses the shared pre-discussion responses of the three models to 200 questions.

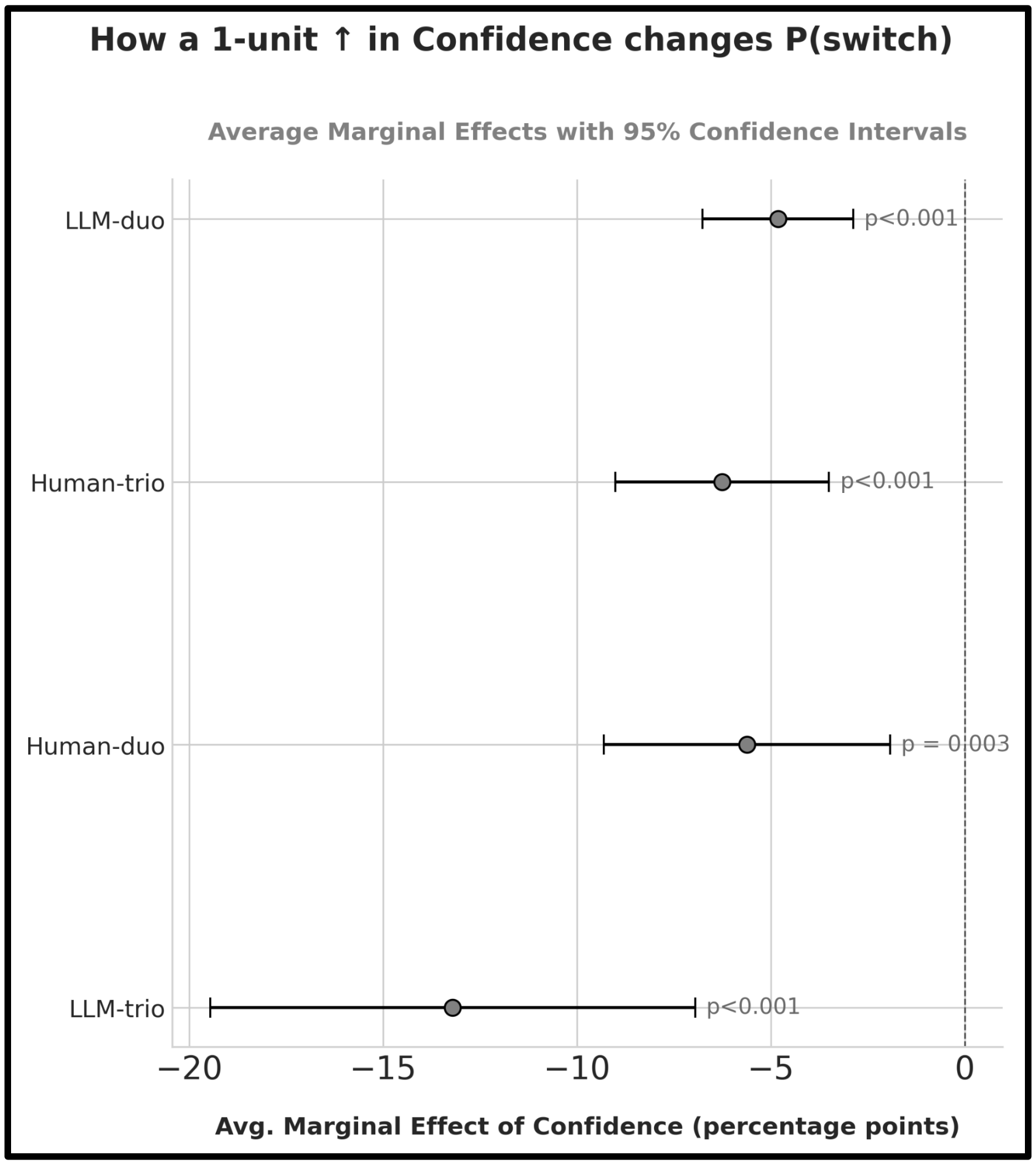

Figure S7. **Higher Confidence Consistently Reduces the Probability of Switching Answers Across All Conditions**. The plot shows the average marginal effect (AME) of an agent's self-confidence on their probability of switching an answer after a discussion, conditional on initial disagreement. Effects were estimated using a logistic regression model for each of the four conditions, which controlled for whether at least one partner had a correct pre-discussion response. The human-trio analysis includes switching outcomes from retained discussion participants only, with all three members supplying partner-answer information. The LLM-pair analysis pools model observations from Experiments 1 and 4. For LLMs in Experiments 1–2, this indicator was positive when at least one of that partner's five sampled responses was correct. The dots represent the AME point estimate (in percentage points) for a one-unit increase in confidence (1–5 for humans and the LLMs in Experiments 1–2; 0–5 for the LLM in Experiment 4), while horizontal bars denote 95% Wald confidence intervals. The consistently negative values demonstrate that higher initial confidence acts as a brake on changing one's mind, after controlling for the partner-correctness indicator. This effect is statistically significant across all four conditions (all $p < 0.01$); the largest absolute point estimate is in LLM trios.

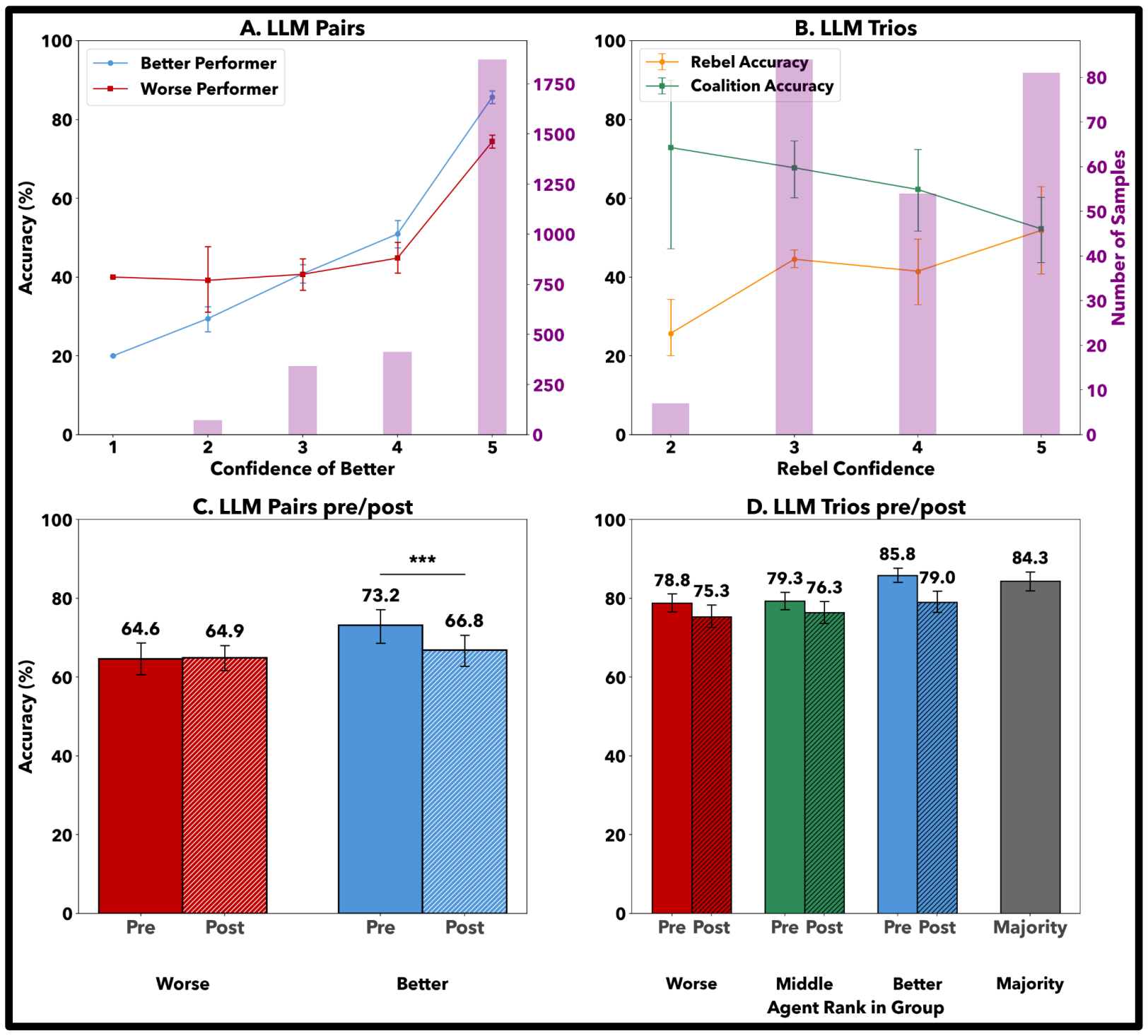

***Figure S8**. LLM-only interactions: Pairs and Trios.* **Replication of our figure 2 results using mid-tier LLMs - Anthropic Claude 3.5 Haiku, OpenAI GPT-3.5, and Google Gemini 1.5 Flash.**

**A.** Knowledge-certainty diversity in LLM-LLM pairs. Accuracy of the better (blue) and worse (red) agents is plotted against the better agent's confidence. Purple bars represent item counts. Calculated Potential Conversation Synergy = 0.3 pp. B. Knowledge-certainty diversity in LLM trios. Rebel (orange) and coalition (green) accuracy is plotted by rebel confidence. Rebel accuracy never exceeds coalition accuracy at any level, reflecting an absence of diversity and explaining the lack of group benefit. Potential Conversation Synergy = 0.0 percentage points. **C.** Pre- and post-conversation accuracy for LLM-LLM pairs in the mid-tier replication of the Experiment 1 protocol, shown separately for the initially better (blue) and worse (red) agents. The worse agent changes little (64.6% → 64.9%); the better agent declines (73.2% → 66.8%). **D.** Pre- and post-conversation accuracy for LLM trios in the mid-tier replication of the Experiment 2 protocol, shown separately by initial performance rank (best, middle, worst) and including the pre-discussion majority vote. All three ranks show accuracy declines following discussion. Error bars in A, B, and D are 95% percentile bootstrap intervals over items (2,000 resamples). Panel B includes only 2–1 split items; PCS retains all 900 trio-items in its denominator. Panel C uses 95% percentile intervals from resampling the nine model-pair-by-benchmark units (2,000 resamples); significance markers indicate two-sided paired t-tests across those units.

This replication was extended to 900 MCQ questions due to affordability of the models. Datasets were slightly adjusted to fit LLMs abilities - we sampled 300 random questions from USMLE-MedQA and MedMCQA and added 300 additional questions from medical topics* from MMLU.

(*) "anatomy", "clinical_knowledge", "college_medicine", "human_aging", "medical_genetics", "nutrition", "professional_medicine", "virology"

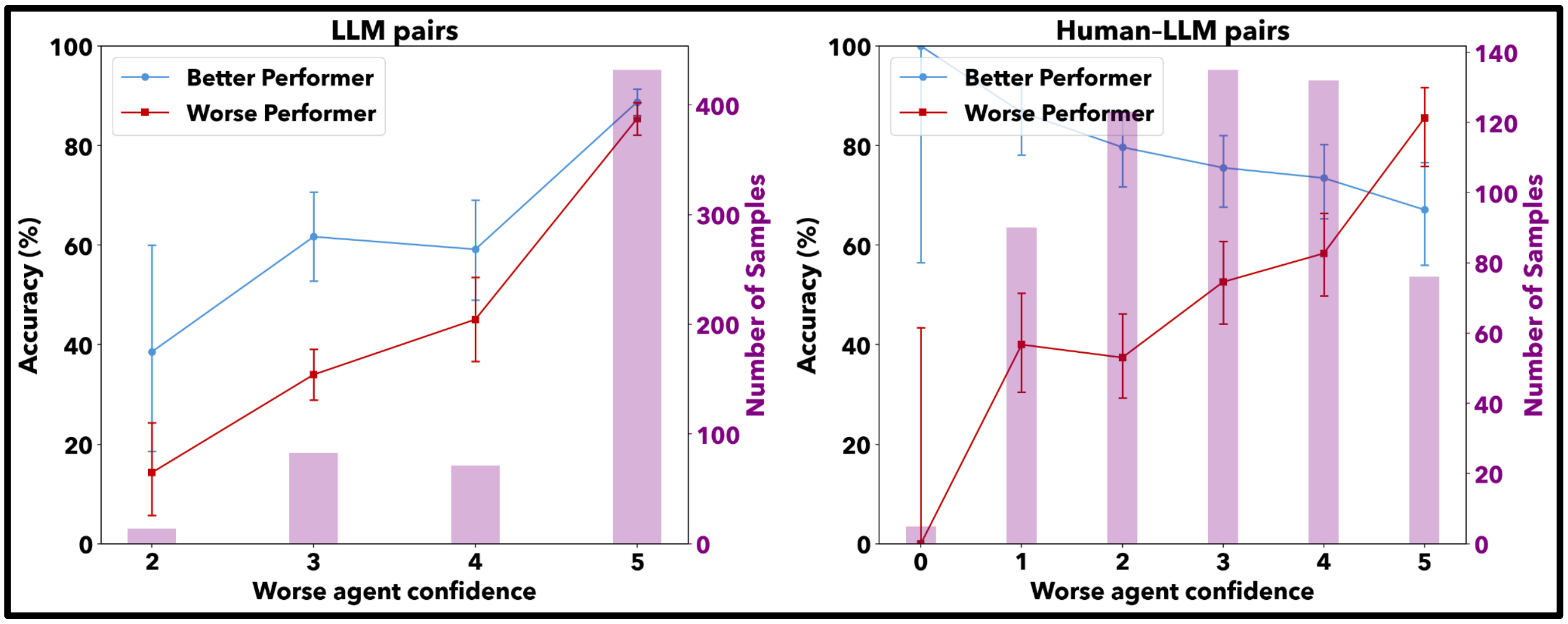

**<u>Figure S9</u>. Diversity modeling using worse agent confidence as the X-axis.**
Accuracy of the better (blue) and worse (red) agents is plotted against the worse agent's confidence. Purple bars indicate item counts. PCS retains the better agent as its accuracy reference. Error bars are 95% percentile bootstrap intervals over items for LLM pairs and 95% Wilson intervals for human–LLM pairs.

**Left**: Knowledge-certainty diversity in LLM-LLM pairs. The computed Potential Conversation Synergy = 0.00 pp.

**Right**: Knowledge-certainty diversity in LLM-Human pairs. The computed Potential Conversation Synergy = 2.50 percentage points.

| Measure | No LLM answer | LLM answer shown | Discussed | p vs. answer shown | p vs. no answer |
|---|---|---|---|---|---|
| Participant–question instances | 214 (20 participants) | 774 (31 participants) | 562 (30 participants) | — | — |
| Initial participant confidence (1–5), mean | 4.22 | 3.13 | 2.86 | 0.006 | < 0.001 |
| Initial participant accuracy | 89.7% | 76.7% | 53.7% | < 0.001 | < 0.001 |
| Mean cohort accuracy on the question | 78.0% | 71.9% | 64.9% | < 0.001 | < 0.001 |
| Initial LLM accuracy (valid answers) | — | 90.1% (n = 768) | 75.3% (n = 546) | < 0.001 | — |
| Initial disagreement (valid answers) | — | 23.7% (n = 768) | 63.9% (n = 546) | < 0.001 | — |

**Table S10**. Initial characteristics by interaction category in Experiment 4. The first two categories had no qualifying discussion. No LLM answer means the LLM never spoke; LLM answer shown means an opening answer was shown without a qualifying discussion. Values are descriptive; p-values compare discussed instances with each non-discussed category using regression of each measure on group, with participant-clustered standard errors. LLM accuracy and disagreement use only valid recorded answers; unavailable answers are missing.

# Appendix

Internal medicine questions used for Experiment 3 and pediatric questions used for Experiment 4 are publicly available in the Hugging Face datasets linked below. The internal-medicine set contains the 41 text-based items. The remaining item used in Experiment 3 (question 25) requires interpretation of two ECG tracings and was omitted from the text-only set used with the language models.

*tomshe/Internal_Medicine_questions_binary* **||** *tomshe/Pediatrics_questions*

**Appendix 2. LLM configurations (LLM-LLM and Trio-LLM)**

LLMs experiment conducted using a version based on Synchronous and Asynchronous User-Customizable Environment for Multi-Agent LLM Interaction [67] available here: https://github.com/Deep-Cognition-Lab/SAUCE

**LLM configuration:**

Models Used  Main Experiments

- **David: Claude 4 Opus (`claude-opus-4-20250514`)**
- **John: o3 (`o3-2025-04-16`)**
- **Alice: Gemini 2.5 Pro (`gemini-2.5-pro-preview-05-06`)**

---

## Prompting Configuration by Stage

Question Prompt presented before conversation for confidence extraction and post-conversation:

***"The following is a multiple-choice question in medicine. Think about it step by step and select the most appropriate option. Format your answer as: The Answer is (X)."***

1. Pre-Discussion (Solo Answering Phase for confidence extraction)

System Prompt:

***"You are a helpful assistant in the medical field, skilled in answering and discussing questions."***

---

**2. Discussion Phase (Conversation Mode)**

**System + Instruction Prompt (for each model):**

***Your name is [Name]. You are a helpful assistant in the medical field, skilled in answering and discussing questions. Your role is to engage in a multi-agent debate about the following multiple-choice question in medicine. Provide your suggested answer, and critically evaluate the answer suggested by another participant. If you believe their answer is incorrect, clearly state why and provide evidence or reasoning to support your claim: [question text]***

---

## Generation Parameters

- Claude and Gemini: Temperature = 0.8, max tokens = 750
- o3: reasoning_effort="high", max_completion_tokens = 10,000

- All models: 1 response per turn (no sampling across multiple completions)

---

## Replication Setup (Figure S8)

For the replication experiment using mid-tier LLMs, the following models were used:

- **David:** Claude 3.5 Haiku **(`claude-3-5-haiku-20241022`)**

- **John:** GPT-3.5 Turbo **(`gpt-3.5-turbo`)**

- **Alice:** Gemini 1.5 Flash **(`gemini-1.5-flash-002`)**

**Answer extraction:**

We used the **EleutherAI/lm-evaluation-harness** multiple-choice extraction library v0.4.0, available at: [https://github.com/EleutherAI/lm-evaluation-harness], with customized regular expressions tailored to capture over 99% of generated response formats.

## Appendix 3. LLM confidence extraction

In Experiments 1–2, each model was run five times using the prompt and configuration detailed in Appendix 2. To reduce positional bias and mitigate memorization effects based on choice order, the answer options were randomly shuffled in each repetition. Confidence was then estimated by computing the frequency of the most commonly selected textual answer (rather than its letter label, due to shuffling), effectively capturing the model's self-consistency across runs. In Experiment 4, confidence instead counts how many of ten independent responses support the model's opening answer; this count is halved and rounded up to give a 0–5 score.

**Temperature checks**: For Claude 4 and Gemini 2.5 Pro we repeated the confidence extraction at temperatures 0.2 and 0.5. For OpenAI we used GPT-4.1, which allows temperature control, under the same settings. Calibration, item level rank correlations, Potential Conversation Synergy ordering, and the ability of Potential Conversation Synergy to predict conversational gains were consistent across temperatures (Fig. S1, S2, S3, S4).

## Appendix 4. Human-LLM Experiment Configuration

We used GPT-4 Turbo, accessed via a custom Telegram client designed to capture conversations and control both the environment and the agent. The Telegram client reset the conversation history whenever the user triggered a predefined signal to initiate a new question for discussion. The model operated using the following prompts, specifically engineered for the conversational setting:

**Initial prompt:**

**sys_prompt =** ***"you are writing a text message to a student who is your colleague, discussing the following question: {question}. Please keep your answers relatively short but explain yourself. provide answers only in Hebrew. Make sure your answer is related to what the user wrote in his last message for example if the user wrote that he thinks that the answer is ב you either explain why it is true if you think it's right or explain why it is wrong if you think otherwise. if he ask what do you think about one of the options you give your opinion and don't hesitate to correct him if he is wrong, he might be confused"***

**In conversation prompt:**
**sys_prompt =** ***"you are writing a text message to a student who is your colleague, discussing the following question: {question}. Please keep your answers short as possible. provide answers only in Hebrew and explain your reasoning a little, no need to call the*** ***person by its name more than once. Make sure your answer is related to what the user wrote in his last message."***

The post-conversation model answer was extracted using the exact same prompts as in the LLM-LLM configurations (Appendix 2): the System + Instruction Prompt from the Discussion Phase, along with the Question Prompt presented after incorporating all user and assistant messages.

## Appendix 5. User questionnaire interfaces

All instructions were presented in Hebrew and are shown here in translation

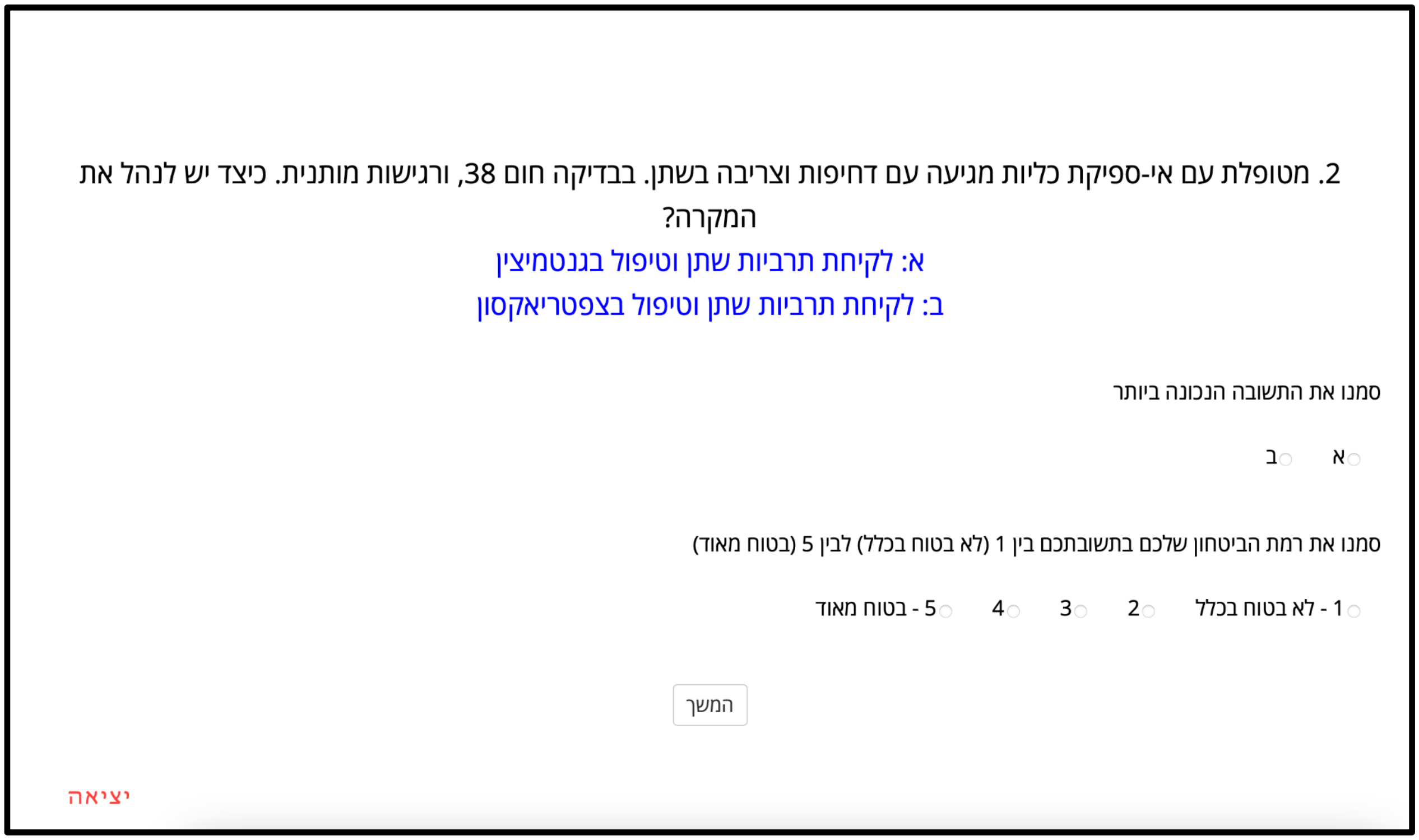

***Figure S10***. **Example of a question interface from the behavioral experiments (3+4).** Participants are presented with a clinical vignette followed by two (experiment 3) or four (experiment 4) answer choices, each corresponding to a different option. Below the answer options, participants are required to select the option they believe is most correct. In addition, they indicate their confidence level in the selected answer on a 1-5 scale, where 1 = "Not confident at all" and 5 = "Very confident". A “Continue” button is used to proceed to the next question, and participants are unable to return to previous questions once they move forward.

# Human-Trio Experiment Instructions

**Stage 1 (Individual Task) Instructions:**

**In the first part of the experiment, you will be required to answer multiple-choice questions with two options in the field of internal medicine.**
The test consists of 42 questions and is allotted approximately one hour to complete. You will not be able to go back and change your answer to a question after proceeding to the next one.
You must answer the questions without using any aids, including internet searches or consulting with others.
You must select an answer for each question in order to move on to the next one.
Additionally, for each question, you will be asked to indicate your level of confidence in your answer.

**Stage 2 (Group Discussion Task) Instructions:**

**In the second part of the experiment, you will be required to answer the same questions from the first part again.**
On the next page, you will find a link to a WhatsApp group. Other participants in the experiment will be added to the group. You may freely discuss the test questions within the group before answering each question.

To keep the discussion organized, you are required to go through the questions in the order they appear. One participant should write the question number (in the following format: **Question Number**) before moving on to the next question.

You are required to answer all questions again, but you are not required to reach agreement on the answer to each question — you should submit the answer you personally believe is most correct.

**<u>WhatsApp Group Rules:</u>**

**In the second part of the experiment, you will be required to answer the same multiple-choice questions again, each with two options, in the field of internal medicine.**
The test consists of 42 questions and is allotted approximately one hour to complete. You will not be able to go back and change your answer to a question after moving on to the next one.
You must answer the questions without using any aids, including internet searches or consulting with others, except for discussions within the designated group.
You must select an answer for each question in order to proceed to the next one.
Additionally, for each question, you will be asked to indicate your level of confidence in your answer.

Maintain an organized discussion. At the beginning of each question, make sure that one of the participants writes the transition to the next question in the following format:
Question [number] in a separate message (e.g., Question 1).

Avoid using the reply feature (Reply) in WhatsApp and do not use emojis during the discussion.

Throughout the conversation, you are asked to state your level of certainty when presenting your opinion and chosen answer.

# Human-LLM Experiment Instructions

**<u>Stage 1 (Individual Task) Instructions:</u>**
**In the first part of the experiment, you will be required to answer multiple-choice questions with four options in the field of pediatrics.**
The test consists of 50 questions and you will have only one hour to complete it. You will not be able to go back and change your answer to a question after proceeding to the next one.
You must answer the questions without using any aids, including internet searches or consulting with others.
You must select an answer for each question in order to move on to the next one.
All questions must be answered within one hour from the start of the test.
Additionally, for each question, you will be asked to indicate your level of confidence in your answer.

**Stage 2 (Bot Discussion Task) Instructions:**
**In the second part of the experiment, you will be able to use a chatbot designed to help you solve the test questions.**
You can access it via the Telegram link provided in the initial experiment message. Before you begin, you must provide the bot with the code: **"medcom"**. You must tell the bot which question you are referring to by writing “Question x” in order to discuss question x. You are required to follow this instruction when moving between questions.

The conversation with the bot is open and unrestricted. If the bot asks for the content of the question, you can simply write the prompt to move to the question again (“Question x”).
You are required to answer all questions again and are not required to agree with the bot on each answer—you should submit the answer you personally believe is most correct.

You are not obligated to discuss every question with the bot, but you cannot go back to a previous question in the test form, and you must complete the test within one hour.

You should begin the second part approximately one hour after the start of the experiment. Please avoid closing the experiment window until the end of this part and the conclusion of the experiment.

**Stage 2 Instructions:**

In the second part of the experiment, you will be required to answer the same multiple-choice questions again, each with four options, in the field of pediatrics.
The test consists of 50 questions, and you will have one hour to complete it.
You will not be able to go back and change your answer to a question after moving on to the next one.
You must answer the questions without using any aids, including internet searches or consulting with others, except for discussions within the designated group.
You must select an answer for each question in order to proceed to the next one.
Additionally, for each question, you will be asked to indicate your level of confidence in your answer.

## Appendix 6. Example of LLM-Human conversations

2 selected examples of LLM-Human discussions translated to English from Hebrew:

---

Q: A 3-month-old girl presents to the emergency department during winter with a runny nose, fever, and breathing difficulties. Examination shows 85% saturation and widespread wheezing and crackles on auscultation. What treatment is required?

Options:

- (A) Oxygen, steroids, and Ventolin inhalation
- (B) Oxygen, steroids, and azithromycin
- (C) Oxygen and GC
- (D) Oxygen

**Discussion:**

**Bot**: "Regarding question 22, the required treatment is (A) Oxygen, steroids, and Ventolin inhalation. The infant has symptoms of breathing difficulty and low saturation, suggesting airway obstruction and inflammation. Ventolin helps expand the airways, and steroids reduce inflammation."

**User**: "You're wrong."

**Bot**: "I apologize for the confusion. Could you share your reasoning? It would help me better understand your perspective."

**User**: "Viral bronchiolitis requires supportive treatment."

**Bot**: "Thank you for clarifying. Indeed, in viral bronchiolitis, the mainstay of treatment is supportive, primarily oxygen. The correct answer here seems to be (D) Oxygen alone, as the goal is supportive respiratory care until the virus resolves spontaneously. Ventolin and steroids haven't shown efficacy in viral bronchiolitis in infants, and thus aren't recommended initially. I hope this clarifies the issue."

**User**: "What is GC?"

**Bot**: "GC stands for glucocorticoids, a type of steroid used for reducing inflammation and managing acute immune responses. In bronchiolitis, they might reduce airway inflammation, but their use should be guided by clinical guidelines and patient response."

---

Q: A child comes to the emergency room with neck stiffness and unilateral neurological symptoms on one side of the face (deviated eye, unilateral facial weakness). What is the correct order of actions?

Options:

- (A) Take blood cultures, perform eye fundus examination and if normal, then perform lumbar puncture followed by antibiotic treatment.
- (B) Take blood cultures, administer antibiotic treatment then perform head CT.
- (C) Take blood cultures, perform head CT and if normal, then lumbar puncture followed by antibiotic treatment.
- (D) Take blood cultures followed by a CTA head urgently.

**Discussion**:

**Bot**: "Regarding question 14, the scenario raises concern for encephalitis or severe neurological conditions. The best initial step is a head CT to rule out complications like bleeding or tumors. If the CT is normal, proceed with a lumbar puncture (LP) and antibiotics. Therefore, the correct answer is (C): take blood cultures, perform a head CT, and if normal, perform LP and then antibiotic treatment."

**User**: "You're wrong; antibiotics should start earlier."

**Bot**: "I understand your point about early antibiotic initiation for suspected encephalitis, but in this case, with symptoms like neck stiffness and unilateral neurological signs, it's crucial to first exclude other conditions such as bleeding or tumors using CT. Doing the CT first prevents incorrect treatment and allows precise and safe therapy once the diagnosis is clearer. Antibiotics can begin immediately after a clear CT result, followed by LP."

**User**: "So what do you think is correct?"

**Bot**: "Considering our discussion and your points, I still believe performing a head CT first is necessary to exclude conditions like bleeding or tumors. Antibiotics and LP should commence once the CT is confirmed normal. This ensures precise and safe treatment, making option (C) most appropriate."

**User**: "I disagree. Neck stiffness and symptoms highly suggest meningitis, requiring early treatment."

**Bot**: "I understand the urgency of treating suspected meningitis promptly. You're correct; antibiotics should often start immediately. However, in this specific case, the unilateral neurological signs necessitate ruling out other serious conditions first via CT. Delaying antibiotics slightly until after the CT doesn't significantly increase risk and ensures accurate treatment. Once the CT is clear, antibiotic therapy and LP should proceed promptly. This approach balances swift action with safety."

**User**: "Therefore, answer (B)."

**Bot**: "I understand your urgency about meningitis. Considering this high suspicion, option (B)—blood cultures, immediate antibiotics, then CT—is indeed appropriate. Early antibiotic initiation is crucial to controlling infection quickly, preventing complications. Thanks for highlighting this important point."